# COVID-19 and Digital Resilience: Evidence from Uber Eats[*†]

Manav Raj,[‡] Arun Sundararajan,[§] Calum You.[**]

**This draft: January 18, 2023**

**Abstract**

Using order-level data from Uber Technologies, we study how the COVID-19 pandemic and the ensuing shutdown of businesses in the United States in 2020 affected small business restaurant supply and demand on the Uber Eats platform. We find evidence that small restaurants experience significant increases in activity on the platform following the closure of the dine-in channel. We document how locality- and restaurant-specific characteristics moderate the size of the increase in activity through the digital channel and explain how these increases may be due to both demand- and supply-side shock. We observe an increase in the intensity of competitive effects following the economic shock and show that growth in the number of providers on a platform induces both market expansion and heightened inter-provider competition. Higher platform activity in response to the shock does not only have short-run implications: restaurants with larger demand shocks had a higher on-platform survival rate one year after the lockdown, suggesting that the platform channel contributes towards long-run resilience following a crisis. Our findings document the heterogeneous effects of platforms during the pandemic, underscore the critical role that digital technologies play in enabling business resilience in the economy, and provide insight into how platforms can manage competing incentives when balancing market expansion and growth goals with the competitive interests of their incumbent providers.

[*] The authors are listed alphabetically by last name. This study was conducted as an independent research collaboration between the authors. Manav Raj and Arun Sundararajan are not and have not been affiliated with Uber Technologies, and Calum You is not and has not been affiliated with NYU. No consulting fees, research grants, or other payments have been made by Uber Technologies to the NYU authors, or by NYU to the Uber Technologies author. We thank Arslan Aziz, Emilie Boman, Libby Mishkin, Sabrina Ross, DJ Wu, Allison Wylie, a set of anonymous editors and referees, participants at the 2020 Workshop on Information Systems and Economics, the 2022 NYC Ops Day, and the NYU Stern Future of Work seminar series for thoughtful feedback, and Uber Technologies for access to data. All errors are our own.

[†] Author emails: mraj@stern.nyu.edu; digitalarun@nyu.edu; calum@uber.com.

[‡] The Wharton School, University of Pennsylvania

[§] NYU Stern School of Business

[**] Uber Technologies

## 1. INTRODUCTION

The COVID-19 pandemic acutely impacted the restaurant industry. As localities imposed shelter-in-place orders and consumers pulled back their presence in public spaces in the first half of 2020, restaurants that relied exclusively or primarily on dine-in faced unprecedented challenges.[1] Numbers from the Advance Monthly Retail Trade Survey by the U.S. Census Bureau indicated an immediate 48.7% year-over-year decline for food services and drinking places in April 2020, reflecting revenue losses of over $30 billion that month alone, and of over $50 billion in March and April 2020 (US Census Bureau Monthly & Annual Retail Trade, 2020). According to the National Restaurant Association, the average establishment saw a 78% decline in its revenues in the first 10 days of April 2020 compared to the same period in April 2019 (Grindy, 2020).

While larger restaurant chains had the resources to weather a prolonged economic downturn, independent establishments and mom-and-pop restaurants struggled to stay alive. Early estimates suggested that up to 75% of independent restaurants would not survive (Severson and Yaffe-Bellany, 2020), as the COVID-19 shock exacerbated credit constraints (Evans and Jovanovic, 1989) that typically disadvantage small businesses which have limited access to alternative financing options (Dietrich, Schneider, and Stocks, 2020). Facing shutdown, restaurants were forced to utilize alternative channels to reach customers, maintain revenue, and survive. A bright spot for some restaurants was the growth in access to digital platforms like Grubhub, DoorDash and Uber Eats in the years preceding COVID-19.

Our study explores the economic effects of access to these channels during the onset of the COVID-19 pandemic in March and April 2020, examining the stabilizing and sustaining effects of a food delivery platform in the weeks following the economic shutdown in the United States and the long-run impact of these effects on the resilience and survival of restaurants. Our analysis also provides new insight into the effects of inter-provider competition on platforms and the effects of pursuing market expansion by adding new providers.

Our analysis is the first to empirically resolve the net impact of potentially countervailing supply and demand aspects of the COVID-induced economic shock, which, a priori, appeared as if they might collectively either raise or lower online performance. The absence of dine-in options may shift demand towards the online channel; variability in the availability of groceries and fears of going to a grocery store may induce consumers to order in rather than cook themselves; and an increase in the hours people have been spending at home coupled with health concerns associated with food prepared and delivered by people outside their homes could induce an increased propensity and ability to prepare one's own meals.

[1] *Appendix Figure A1* presents the year-over-year percent change in seated diners from mid-February to mid-April 2020.

We use order-level data from the Uber Eats food delivery platform for five major US cities (New York City, San Francisco, Atlanta, Miami, and Dallas) from February 1st through May 1st, 2020, to examine changes in digital demand and performance following the shutdown. We show that the net *demand-side shock* was consistently positive but varied substantially in magnitude across geographies despite the likely negative demand spillovers from a *supply-side shock* induced by the shuttering of many restaurants and the cutting back of operating hours for those that remained open. Across our five sample cities, we see that restaurants that remained open for delivery experienced significant and economically meaningful increases in the count of orders that they received. We show that the size of the demand shock experienced by restaurants during this period is moderated by both restaurant- and locality-specific characteristics. For example, higher-rated restaurants and restaurants that have spent more time on the platform experience a larger increase in orders during the post-period. The increase in orders for restaurants that remained open on the platform occurred despite many of these restaurants cutting the number of hours that they offered supply on the platform but was perhaps aided by other restaurants shutting down their digital channel entirely.

We contribute new findings to a growing literature that has explored the challenges and opportunities created by digital distribution channels for incumbent firms (e.g., Chen, Hu, and Smith, 2019; Danaher *et al.*, 2010; Smith and Telang, 2009), and in particular, smaller businesses. For example, Airbnb is now a channel not just for sharing one's spare bedroom or apartment, but also for incumbent bed-and-breakfasts and full-time independent vacation rental properties (Guttentag, 2015). Similarly, e-commerce platforms such as eBay and Amazon.com have long been critical channels for small businesses and individual sellers (Bailey *et al.*, 2008). Our study underscores how COVID-19 has now made food delivery platforms vital to small businesses in the restaurant sector. Further, we contribute to a body of work that explores how the net economic effects of a platform channel can vary depending on establishment or locality characteristics (e.g., Brynjolfsson, Hu, and Simester, 2011; Dewan and Ramaprasad, 2012; Raj and Eggers, 2022).

We further explore the competitive implications of the supply-side shock induced by the shuttering of restaurants by examining how competitive intensity moderated the demand shock experienced by restaurants following the closure of the dine-in channel. We show that the intensity of inter-provider (restaurant) competition increased during the COVID-19 shock. We do so by constructing a new and more economically robust metric of competition based on actual overlapping consumption across businesses and inferred from a bipartite graph (e.g., Bimpikis, Ehsani, and İlkılıç, 2019; Huang, Zeng, and Chen, 2007; Leiponen, 2008) of customers and restaurants. This novel approach may independently contribute towards better measurement of platform-based inter-provider competition in future work. As one might expect, the performance of an individual restaurant on the Uber Eats platform is better when it faces less competition, with a larger effect of competition in the post-shock period. By probing deeper into the competitive effects of this supply-side shock, we add to prior work that has examined how digital platforms alter the nature of competition (Animesh, Viswanathan, and Agarwal, 2010; Farronato and Fradkin, 2018; Filippas, Horton, and Zeckhauser, 2020; Zervas, Proserpio, and Byers, 2017). These divergent effects of market expansion and inter-provider substitution provide new insight into the tension between what is optimal for a platform and what is optimal for its complementors.

Finally, we examine platform contributions to longer-term resilience, showing that the magnitude of the demand shock experienced during the start of the pandemic is related to whether the restaurant remained on the platform over the next year. Our findings suggests that the platform was a source of digital resilience for restaurants that were able to take advantage of the opportunity that the digital channel offered during this period of crisis. By illustrating how and when digital channels can stabilize firm performance when other distribution channels are limited, we add a new facet to a body of research on cannibalization and expansion due to digital distribution channels (Chen *et al.*, 2019; Danaher *et al.*, 2010; Smith and Telang, 2009; Viswanathan, 2005; Yin *et al.*, 2009) and on the interplay between platform-mediated and physical-world opportunities for small businesses (Han *et al.*, 2019; Kitchens, Kumar, and Pathak, 2018).

Our investigation of the moderating effect of competition on performance on a digital platform adds to a body of literature that explores how digital technologies and distribution channels affect substitution and spillovers across competing firms (Haviv, Huang, and Li, 2020; Liang, Shi, and Raghu, 2019; Raj, 2021; Reshef, 2022; Tambe and Hitt, 2014). In addition, our analysis of the dual effects of market expansion and business stealing contributes to literature that discusses the tension between strategies that are optimal for the platform and strategies that are optimal for platform participants, complementors, and co-creators (e.g., Castillo, 2020; Ceccagnoli *et al.*, 2012; Filippas, Jagabathula, and Sundararajan, 2021; Rietveld, Ploog, and Nieborg, 2020). By identifying which restaurants were best able to take advantage of the platform during the early period of the pandemic, our work further adds to new line of literature that seeks to understand how the COVID-19 pandemic affected small businesses (de Vaan *et al.*, 2020) and online channels (Han *et al.*, 2020). Further, our work presents practical implications for small businesses by highlighting how digital distribution channels can serve as a differentiating source of resilience during crisis.

The rest of this paper is organized as follows. Section 2 describes our empirical setting and data and presents model-free evidence. Section 3 outlines our baseline empirical models and present evidence of the average change in activity on the platform following the imposition of shelter-in-place orders across the sample cities. Section 4 analyzes how changes in peer restaurant supply on the platform shapes the demand shock that a specific restaurant experiences. Section 5 examines restaurant- and locality-specific drivers of heterogeneity in the demand shock experienced. Section 6 explores the relationship between the increase in activity on the platform during this period of crisis and restaurant survival on the platform over a longer period of time. Section 7 concludes and outlines directions for additional inquiry.

## 2. EMPIRICAL SETTING AND DATA

As the COVID-19 virus began to spread across the United States in March 2020, states and cities requested or required that residents avoid any non-essential travel or activity (Mervosh, Lu, and Swales, 2020). Such "stay-at-home" or "shelter-in-place" orders, in conjunction with consumer anxiety and fear regarding the virus, devastated economies, leading to rising unemployment rates, falling consumer activity, and stifled growth (Lee, 2020). We use the economic lockdowns associated with the COVID-19 pandemic in March and April 202 as an exogenous shock to study how digital channels can contribute to short-run and long-run economic value when traditional channels are disrupted.

To study how the onset of this crisis affected restaurant performance on digital channels, we rely on data from the Uber Eats platform, a digital food delivery platform offered by the ridesharing company Uber. On Uber Eats, consumers can review menus and order food for delivery or takeout from participating restaurants using an application provided by the platform or through a web browser. In exchange for hosting the transaction and connecting consumers to restaurants, Uber Eats collects a commission on the orders placed on the platform from restaurants and collects delivery charges from customers.

We use order-level data to examine how restaurant demand and performance on the Uber Eats platform changes following the shutdown. Our sample consists of activity on all "Small or Medium Business" (SMB) restaurants, defined as restaurant chains with fifty or fewer locations on the Uber Eats platform, across five major US cities (New York City, San Francisco, Atlanta, Miami, and Dallas) from February 1 to May 1, 2020. In a subsequent part of the paper, we also examine 2021 data to assess restaurant survival. A vast majority of these restaurants (over 90% of the SMBs in our sample) have just a single location.

To identify the effect of the closure of dine-in restaurants on restaurant performance on the Uber Eats platform, we take advantage of the enactment of shelter-in-place guidance in each of our five sample cities. For each city, we define a "pre-lockdown" and "post-lockdown" period depending on the date the city enacted shelter-in-place guidance (for all cities, this occurred in mid-March 2020). For example, for New York City, the pre-shutdown period comprises the days February 1 through March 16, and the post-lockdown period comprises the days March 17 through May 1.[2]

[2] *Appendix Table A2* summarizes the dates used to define the pre- and post-lockdown periods for each city. Our results are robust to using March 16th as the cut-off date for the pre- and post-period for all sample cities (see, *Appendix Table A3*).

We measure restaurant demand and performance using the daily count of orders a restaurant receives.[3] For data anonymity purposes and to ease interpretation of coefficients, order counts at the restaurant-day level are scaled by the pre-period mean count of daily orders across all restaurants within the city. This scaling allows us to interpret changes in levels and coefficients as percentage changes in daily orders relative to pre-period activity levels within the city.

In an ideal natural experiment, we would have a set of control cities that were unaffected by the pandemic. If such a set of control cities existed, we could use a difference-in-differences identification strategy to evaluate the causal impact of restaurant closure by comparing trends from the pre-period to the post-period for treatment vs. control cities. Unfortunately, the pandemic affected all major cities worldwide, leaving no appropriate sample of control cities and limiting our ability to make causal inferences. Nevertheless, even while limited to comparing within-city changes pre-lockdown vs. post-lockdown, we believe that the exogenous nature of the pandemic provides insights into how a change in the availability of a traditional channels affects supply and demand through alternative digital channels.

We have considered the possibility that a waiver of delivery fees or changes in commissions by Uber Eats during the lockdown could confound our results.[4] For this reason, we explored regulations that capped commission or delivery fees, towards considering whether these caps were the primary factor in determining any demand shocks we measured. As it turns out, just one of our five sample cities (San Francisco) implemented commission caps during our sample period (on April 13, 2020). Because we are able to document evidence of demand- and supply-side effects across cities, we concluded that the commission cap alone was not driving our results. While we could have learnt more had we been able to acquire data about delivery fees and commission rates charged across time by restaurants in our sample, that data is not available to us. Thus, the results we document are induced not just by the shuttering of physical restaurants, but also by the choices that platforms made during the shock. For example, during the pandemic Uber Eats waived delivery fees for independent restaurants (Korosec, 2020).

We first provide model-free evidence aligned with the demand-side and supply-side shocks discussed in the introductory section. In Figure 1, we document that the imposition of shelter-in-place orders elicited dramatic changes in consumer behavior on alternative distribution channels. Figure 1, Panel A shows how consumer sessions with demand intent, defined as any instance when a consumer opens the application and then takes an action indicative of consideration of purchase (e.g., adding an item to their cart, searching for a specific restaurant, opening a restaurants profile), increased post-shutdown across our five sample cities, with an increase in average daily consumer activity that ranges from 25% in Miami to nearly 75% in San Francisco.

[3] Because of the proprietary nature of the Uber Eats data, we are limited in the outcome measures we can consider at part of this research study, and do not have access to restaurant revenues or order sizes.
[4] We thank an anonymous referee for helpful feedback on this issue.

We next turn to examining how consumption changed pre- and post-intervention. As illustrated in Figure 1, Panel B, following the lockdown, all five cities experienced an increase in the total number of Uber Eats orders placed across all restaurants. The average total number of orders per day was significantly higher in the post-lockdown period across all cities except New York City, with the highest increase in San Francisco (51.9%, $p < 0.01$) and the lowest in New York City (4.1%, $p = 0.13$). This increase in total daily orders across cities occurred despite conversion rates, defined as the percent of consumer sessions with demand intent that end with a completed order, decreasing or staying at similar levels to the pre-period across all sample cities (illustrated in Figure 1, Panel C). This indicates that the increase in orders was largely driven by an increase in consumer activity on the platform rather than through greater conversion of existing visitors or traffic.

We examine the extent to which orders were made by new (no orders placed prior to the sample), infrequent (less than five orders in the three months prior to the sample), and frequent consumers (five or more orders placed in the three months prior to the sample) to explore whether the increase in demand was driven by new vs. regular consumers. During the pre-period of the sample, 60.4% of orders were placed by frequent consumers, 23.7% of orders were made by infrequent consumers, and 15.9% of orders were placed by new consumers. In comparison, during the post-period of the sample, 44.5% of orders were placed by new users, 25.7% of orders were placed by infrequent consumers, and 29.8% of orders were placed by new consumers. This suggests that the increase in activity on the platform was at least partially driven by an increase in activity by new consumers.

The increases in orders on the platform occurred despite widespread restaurant shutdowns and cutbacks on operating hours. Many restaurants shuttered entirely, either temporarily or permanently, and of those that did not, many cut back on the hours of operation because of the dramatic decrease in on-site dining activity. As illustrated in Figure 2, following the lockdown, all five cities witnessed a substantial reduction in both the number of restaurants open for business on the platform (Panel A) and the average number of supply hours (Panel B).[5] This decrease in supply may explain why conversion rates fell across sample cities, as consumers may have found that many of their favorite restaurants were no longer accepting orders.

The decrease in the number of restaurants open for delivery was largest in New York City (38.1%, $p < 0.01$) and smallest in San Francisco (7.6%, $p < 0.01$). The decrease in the number of supply hours from restaurants that remained open was fairly consistent across cities (ranging from 6.6% to 9.1%). Thus, the net drop in supply is quite profound across cities, and most dramatic in New York City, where total daily supply hours post-shutdown was, on average, barely half the pre-shutdown level. There are many factors that could have caused this precipitous drop, including restaurants ceasing operations (either temporarily or permanently) or restaurants remaining open but cutting back on operating hours because of the loss of dine-in customers. A comparison between Figure 2, Panel A and Figure 2, Panel B indicates that both these factors are at play. In New York City, the supply squeeze is driven largely by restaurants closing down, while in San Francisco, cutbacks in operating hours accounts for most of the decline in supply.

We explore which factors are correlated with a restaurant's decision to offer supply during the post-period in *Appendix Table A4*. Controlling for either city or zip code fixed effects, we show that the restaurants most likely to offer supply during the post-period are the restaurants that were most active on the platform prior to the shock. Further, restaurants are more likely to offer supply if they have been on the platform for a longer period of time, if they have higher ratings on the platform, have lower prices, and if peer restaurants (as defined using the network-based competition measure described further in the draft) also remain open. Restaurants located in more densely populated areas were less likely to offer supply, perhaps because concerns around the transmission of the virus remained highest in dense urban areas. We return to this analysis later in the paper.

The simultaneous increases in demand and decreases in supply dramatically raised the rate at which active restaurants received and fulfilled orders. As illustrated in Figure 3, order velocity (number of orders fulfilled per available supply hour) increased dramatically across all five cities. The velocity increases are most dramatic in New York City and Dallas, where on average, restaurants that were still operating post-lockdown more than doubled the number of orders they are fulfilling per available hour but are by no means small in other cities: nearly doubling in San Francisco, 74% higher in Miami and 71% higher in Atlanta. We present summary statistics of the relevant variables used in our analyses in Table 1.[6]

[5] A restaurant offering supply hours has indicated to the Uber Eats platform at some point during the day that it was available to take orders. We quantify the windows of time that a restaurant indicates availability in this manner by measuring, for each restaurant on each day, its number of supply hours.

Because we do not have data on dine-in sales during the pre- or post-period, we cannot say whether the demand shocks documented on the platform represent cross-channel cannibalization or increases in total demand. Given that in-person dining was fairly low during the shelter-in-place orders, it is likely that the demand spillovers documented are more likely to represent cross-channel cannibalization. However, even if this is the case, we believe this represents important evidence that the platform provided a compensatory channel for the restaurants during the crisis.

## 3. BASELINE MODEL: IMPACT OF LOCKDOWN ON ORDERS

Our model-free analysis reveals an increase in total transaction activity coupled with a sharp decline in supply. The analysis that follows aims to better understand restaurant-level impacts, towards unpacking the effects that the availability of the digital channel has on SMB restaurants. We quantify this impact by focusing on one key dependent variable – *the daily number of orders received by a restaurant.* While acknowledging that daily revenue would have been an informative metric, confidentiality concerns precluded the analysis of this measure.

Our first model assumes the following form:

$$Y_{it} = \beta_1 Post_{it} + Restaurant\ FE_i + Day\ of\ Week\ FE_t + \varepsilon \qquad (1)$$

In this equation, $i$ indexes the restaurant and $t$ indexes the date. $Y$ is the count of orders a restaurant receives in a day, our dependent variable, and *Post* is the main independent variable which indicates whether the city has enacted a shelter-in-place order on a given day. *Post* is equal to zero on all pre-lockdown days and is equal to one in all post-lockdown days. Our period of analysis does not include any dates when lockdown orders previously imposed have been lifted. To account for restaurant-specific heterogeneity that may drive our result (including restaurant popularity, cuisine, and the choice of whether to remain open for delivery following the shelter-in-place order) we include restaurant fixed effects. We also include day-of-week fixed effects to account for cyclical variation in order density.

[6] Variable definitions can be found in *Appendix Table A1*.

In Table 1, we present results from the regression analysis examining the effect of the shelter-in-place orders on orders on the Uber Eats platform. In Column 1, the sample includes all restaurants that offered some positive level of supply (availability to take orders) on Uber Eats platform in the pre-lockdown or post- lockdown period. In Column 2, we display results for a sample of all restaurant-days for which a restaurant offers supply on the Uber Eats platform.

In both samples, we find that restaurants receive significantly more orders in the post-period than in the pre-period, even controlling for restaurant and day-of-week fixed effects. In Column 1, in the sample of all restaurants that offered supply in either the pre- or post-period, we find that restaurants on average received 12.5% more orders a day in the post-period relative to the pre-period ($p$<0.01). In Column 2, using the sample that only contains restaurant-days for which a restaurant offers supply, this effect is much larger, and we find that restaurants that offer supply on a given day receive 44.6% more orders a day in the post-period ($p < 0.01$). We model how this effect manifests temporally throughout the pre- and post-period for both samples in Table 1 in Figure 4. This figure shows that the increase in orders on the platform following the imposition of shelter-in-place orders began immediately and grew rapidly over the first four weeks following the orders. It remains relatively consistent in weeks five through eight following the shelter-in-place orders. This pace of growth may reflect initial uncertainty regarding how the virus spread leading to cautious and gradual adoption of food delivery in the immediate aftermath of the shelter-in-place guidance.

The difference in effect size across the two samples is because the sample used in Column 1 continues to include, post-lockdown, restaurants that were active prior to the declaration of shelter-in-place guidance but are inactive in the post-period. Thus, a number of restaurants in this sample have zero daily orders for all or most of the post-lockdown time window. We explore the competitive implications of these supply-side changes later in the paper, and in what follows, focus exclusively on the latter sample, since our primary interest is in restaurant-level impacts.[7]

[7] Results are also robust to a sample of restaurants that offered supply prior to the shock (*Appendix Table A5*).

We next consider heterogeneity in this effect across our five cities. We interact the post-shutdown indicator with a binary variable for each city to calculate a city-specific effect. Our model takes the following form:

$$Y_{it} = \beta_1 Post_{it} + \beta_2 Post_{it} \times City_i + Restaurant\ FE_i + Day\ of\ Week\ FE_t + \varepsilon \qquad (2)$$

For this analysis, we focus on restaurants offering supply on a given day and use the sample comprising only restaurant-days for which a restaurant offers supply on the Uber Eats platform (the same data used in Table 2, Column 2). The results of this analysis are presented in Table 2, Column 3.

Combining the baseline post-effect with the interaction effect for each city in Table 2 allows us to separately identify how the average count of orders per restaurant changes across cities in the post-period. Figure 5 illustrates these results of Table 2, column 3, depicting the city-specific marginal effects within each city during the pre-period. A restaurant in San Francisco that continues to offer supply on the Uber Eats platform following shelter-in-place guidance and was getting 10 orders a day on average prior to March 16 would be, on average, receiving 16 orders a day post-shelter-in-place guidance (the marginal effect is a 61.1% increase, significant at $p < 0.01$). Similarly, Dallas restaurants that remained open and available on the platform have experienced average increases of 53.5% and NYC restaurants experienced an effect of about 47%. Restaurants in Miami saw the smallest increase (23.2%); however, the measured effect is statistically and economically significant (at $p < 0.01$) across all cities.

The changes we document in Uber Eats ordering activity may be due to several reasons. The absence of dine-in options may shift demand towards the online channel; variability in the availability of groceries and fears of going to a grocery store may induce consumers to order in rather than cook themselves. In contrast, an increase in the hours people have been spending at home coupled with health concerns associated with food prepared and delivered by people outside their homes could induce an increased propensity and ability to prepare one's own meals; the lack of availability from one's favorite restaurants on the platform could add to this negative demand effect. Our evidence indicates that, in the aggregate, these and other negative demand effects are significantly outweighed by the positive.

## 4. SUPPLY SQUEEZE AND COMPETITIVE EFFECTS

The changes in restaurant supply on the platform is likely to influence the magnitude of the demand shock that a specific restaurant experiences. For example, restaurants may see particularly large increases in demand and orders if their direct competitors are unable to stay open or provide less supply.

We thus investigate how a contraction in competitor supply on a digital platform, such as the one witnessed in our empirical setting, affects demand for peer restaurants and analyze how restaurant supply hours and competitive conditions influence the effect of the COVID-19 induced demand shock. To assess the potential moderating effect of competition in determining whether and how the closure of dine-in restaurants affects restaurant performance eon the Uber Eats platform, we require a measure of competition.

### *4.1 Two measures of competition*

We construct two measures of competition using consumer-level data from Uber Eats. The first measure defines each restaurant's competitive set as the other restaurants within the same primary cuisine categorization within the city that provide supply on the Uber Eats platform at some time between February 1 and May 1, 2020. For this measure, the competition index is the percentage of restaurants within a category offering supply on a given day. A higher value in this measure suggests that a consumer would have more options on the Uber Eats platform to choose from within that cuisine category. We call this the ***shared category*** competition measure.

Our second measure of competition is novel. We use order-level data from Uber Eats in the year prior to the beginning of our sample period (the "training period" February 2019 through January 2020) to construct a bipartite consumer-restaurant graph that is inspired by measures of product complementarity used in the collaborative filtering literature. Let $x_{ij}$ be the number of orders customer $i$ has placed at restaurant $j$ during the training period and let $r_j = \sum_{i=1}^{m} x_{ij}$ and $c_i = \sum_{j=1}^{n} x_{ij}$ be the number of orders received by restaurant $j$ and placed by customer $i$ respectively. Then the competitive intensity of restaurant $b$ on restaurant $a$ is the sum, across all customers of restaurant $a$, the likelihood each customer orders from restaurant $b$ weighted by the importance of the customer to restaurant $a$, or

$$CompetitiveIntensity(a, b) = \sum_{i=1}^{n} \frac{x_{ia} x_{ib}}{r_a c_i}.$$

Breaking this measure down into its component parts, the "importance" of customer *i* to restaurant *a* is simply the count of orders from customer *i* to restaurant *a* divided by the total count of orders received by restaurant *a* during the sample period ($\frac{x_{ia}}{r_a}$). Correspondingly, the likelihood that customer *i* orders from restaurant *b* as the count of orders from customer *i* to restaurant *b* divided by the total count of orders made by customer *i* during the sample period ($\frac{x_{ib}}{c_i}$). Summing this product across all overlapping customers generates the dyad-level competitive intensity score for the pair (*a*, *b*). The level of competition that a restaurant faces on a given day is thus simply the competitive intensity scores for each dyad the restaurant is in for which the competing restaurant is open on that day.[8] We call this the ***customer network*** competition measure.

*4.2 Model and Results*

We use a model of the following form to estimate the role of competition in determining the effect of the COVID shock on restaurant performance:

$$Y_{it} = \beta_1 Post_{it} + \beta_2 Competition\ Index_{it} + \beta_3 Post_{it} \times Competition\ Index_{it} + Restaurant\ FE_i + Day\ of\ Week\ FE_t + \varepsilon \qquad (4)$$

By interacting each of our measures of competition with the post indicator variable, we identify how the level of competition on a given day moderates the effect of dine-in restaurant closure on restaurant performance on the Uber Eats. Comparing the effect of our competition indicator in the pre- vs. post-period, we identify whether and how the relative supply of peer restaurants affects the increase in restaurant demand following the shock. Our results are presented in Table 3.

Columns 1 and 3 indicate that predictably, competition has a negative effect in both the pre- and post-period. The interaction effects between the post-indicator and the two measures of competition reveals how the level of competition and its effect on outcomes evolve following the imposition of the shelter-in-place orders. Specifically, the effects of competition as measured by both the shared category and customer network measures are <u>stronger</u> in the post-period. Column 2 suggests that a 10% increase in the number of restaurants open within a cuisine category is associated with a 6.3% decrease in the count of orders a restaurant receives in the pre-period (at $p < 0.01$), while the equivalent increase in the number of restaurants open within a cuisine category is associated with an 8.6% decrease in the count of orders a restaurant receives (at $p < 0.01$). The difference is starker with the network-based measure, as the estimates in column 4 suggest that this measure had no meaningful relationship with restaurant orders in the pre-period, but post-lockdown, a 10% increase in the network measure is associated with 9.1% fewer orders (at $p < 0.01$). It is possible that, pre-lockdowns, the presence of similar restaurants with overlapping customers may have generated positive spillovers through indirect network effects (Katz and Shapiro, 1985; Rochet and Tirole, 2003; Weyl, 2010). Put differently, pre-COVID, it could be that consumers are attracted to the platform by one restaurant and then explore and discover other related restaurants but following the lockdown and the ensuing increase in demand (and consumer activity), such spillovers no longer stimulate sufficient demand to overcome the negative effects of substitution. We illustrate the marginal effect of a change in competition across both considered measures during the pre- vs. post-period in Figure 6.

[8] *Appendix Figure A2* displays the distribution of the competitive indices pooled across all sample cities for the whole sample, pre-, and post-period.

As shown by the results in Table 3, more competition on the platform decreases the count of orders each restaurant receives. However, what is unfavorable for the restaurants on the platform may be favorable or unfavorable for the platform as a whole. We investigate whether this is the case by exploring whether and how supply on the platform on a given day is associated with the platform's performance, as measured by the total count of orders a platform receives in a locality on a given day. We estimate models of the following form at the ZIP code level:

$$Y_{it} = \beta_1 RestaurantSupply_{it} + Zip\ Code\ FE_i + Date\ FE_t + \varepsilon \qquad (5)$$

In this model, $i$ indexes the ZIP code and $t$ indexes the date. The dependent variable $Y$ is the log-transformed count of orders a received by restaurants located in a ZIP code on a given day, and *Restaurant Supply* is the main independent variable which indicates the supply of restaurants available on the platform on a given day.

We operationalize this supply variable in three ways: (i) the raw count of restaurants located in the ZIP code offering supply; (ii) the log-transformed count of restaurants located in the ZIP code offering supply, and (iii) the percent of restaurants located in the ZIP code offering supply.

To account for location-specific effects, we include ZIP code fixed effects. We also include date fixed effects to account for variation in demand shocks over time.

The results of this analysis are presented in Table 4. We find that across our sample, a greater supply of restaurants in a locality on a given day is associated with greater platform performance. The estimates in column 1 suggest that each additional restaurant offering supply on the platform is associated with a 1.8% increase in total orders received by the platform (at $p < 0.01$). This positive relationship persists when considering the log-transformed number of restaurants offering supply (column 2) as well as the percent of restaurants offering supply (column 3).[9] This positive relationship exists in both the pre- and post-period (columns 4-6 and 7-9 respectively).

Taken together, these results suggest a tension between what is best for the platform and what is best for providers on the platform. Greater supply on the platform benefits the platform by stimulating greater consumer demand. However, greater supply on the platform also increases competitive intensity for each restaurant competing upon them platform, affecting these providers negatively. Balancing these countervailing incentives for the platform and its providers is a critical strategic issue for platforms.

It is possible that, over a longer period of time, these competitive effects may be offset by indirect network effects if a greater number of providers brings more consumers to the platform. However, in the interim, platforms that induce too much of a competitive squeeze in pursuit of market expansion may lose providers to competing platforms or risk having them poached by new entrants, an issue complicated further by the inability of most platforms to maintain exclusive relationship with providers owing to the associated labor regulation risks. A deeper exploration of strategies for navigating this path represents a promising area for future work.

## 5. DRIVERS OF HETEROGENEITY IN DEMAND SHOCKS

Thus far, we have documented how restaurants on the Uber Eats platform experienced a demand shock in the immediate aftermath of the pandemic and that changes in the competitive environment on the platform influenced the magnitude of the shock. We now investigate how selected restaurant- and locality-level characteristics may moderate the effect of an economic shock (in our case, the pandemic shock) on restaurant performance. In doing so, we identify which restaurants were best able to take advantage of the platform during this period of crisis. We estimate models of the following form at the restaurant-level:

$$Y_{it} = \beta_1 Post_{it} + \beta_2 Post_{it} \times Characteristics_{it}$$
$$+ Restaurant\ FE_i + Day\ of\ Week\ FE_t + \varepsilon \qquad (6)$$

[9] This positive relationship also persists when considering the raw count of orders on the platform rather than the log-transformed count of orders as a dependent variable. See *Appendix Table A6*.

By interacting each of the "post" indicator variable with different restaurant and locality-level characteristics, we generate insight into the various factors that moderate the size of the demand-side shock experienced by restaurants on the platform. At the restaurant-level, we investigate the moderating role of restaurant pricing, consumer ratings on the platform, and age; at the locality-level, we consider the moderating role of local average income, technology usage (smartphone and broadband penetration), and population density. We also estimate the role of these restaurant and locality-level characteristics in determining the effect of the COVID shock on restaurant performance by interacting each of the used considered moderators with the post-shock indicator variable. The results of this analysis are in Table 5.

Our results uncover an interesting pattern that identifies restaurants best able to take advantage of the demand-side shock on the Uber Eats platform following the shelter-in-place guidance. We find, for example, that the demand increase was greater for higher-rated restaurants (column 3) and restaurants that had been on the platform longer (column 4). Each additional "star" rating (out of five) is associated with a 16.1% increase in the demand shock experienced, or stated differently, a one standard deviation increase in average rating is associated with a 6.9% increase in the size of the demand shock. Similarly, each additional year on the platform is associated with a 2.3% increase in the size of the demand shock experienced.

Intuitively, this pair of results makes sense because highly rated and older restaurants may have been better positioned to take advantage of the increase in consumer activity on the platform. More positive consumer evaluations may have allowed highly-rated restaurants to attract new consumers, and more experience with the digital channel could lead to a greater familiarity working with the platform and its infrastructure, enabling a restaurant to more quickly transition to conducting its business primarily through the platform-mediated channel.

Next, we turn to locality characteristics. We find that the size of the demand-side shock decreased with average locality household income (column 5) and increased with population density (column 8). A one standard deviation increase in average household income is associated with a 3.5% smaller demand shock, a one standard deviation increase in smartphone penetration is associated with a 2.2% smaller demand shock, and a one standard deviation increase in population density is associated with a 2.2% larger demand shock.[10]

These results may indicate that the presence of a platform like Uber Eats was especially helpful in areas that were particularly vulnerable to the negative effects of the pandemic shock: densely populated urban locations that may have seen particularly stark decreases in activity due to fears of transmission, and lower income areas that faced intense economic uncertainty following the pandemic shelter-in-place guidance.

The latter result is likely partly driven of a high level of usage of the platform in higher income area pre-pandemic. However, it could also be a consequence of a contrast in changes in day-to-day life across lower and higher income neighborhoods. People with higher-income jobs were more likely to work from home, and in cities like New York and San Francisco, more likely to have kitchen space amenable to frequent meal preparation. This is in contrast with people with lower-income jobs, who may have been less likely to have either the leisure time or the infrastructure for in-home meal preparation.[11]

The finding is also important in light of related findings (see, e.g., Chang *et al.*, 2021) that visits to restaurants were among the most salient drivers of the spread of COVID-19 during this period, and that marginal effect of physical restaurant visits was most acute in the lowest income neighborhoods. Our finding thus suggests that the presence of platforms like Uber Eats may have mitigated this extreme adverse effect of restaurant visits, and that the contrasts observed by Change et al. (2021) between lower and higher income neighborhoods may have been even starker were it not for the availability of the digital channel.

[10] We also find that one standard deviation increase in smartphone penetration is associated with a 2.2 percent smaller demand shock, and do not find any evidence of an effect of broadband penetration. We conducted these analyses to explore the moderating effects of technological sophistication, but we have not inferred anything especially important from the regression results.

[11] We thank an anonymous referee for suggesting this direction of analysis and this possibility.

## 6. LONG-RUN EFFECTS OF PLATFORMS ON DIGITAL RESILIENCE

Thus far, we have shown how restaurants benefitted in the short run from the presence of digital channels like Uber Eats, as the platform allowed them to mitigate some of the effects of losing their physical sales channel during the lockdowns. We now turn to examining long-run effects, and specifically exploring how these short-run demand increases that smoothed out the adverse economic effects of shelter-in-place might have altered a restaurant's ability to survive beyond this immediate period of crisis.

To identify whether and to what extent the platform-mediated demand shock served as a source of resilience for restaurants during this period of crisis, we conduct a *survival analysis*. We do this by examining the relationship between the demand shock experienced and the likelihood of restaurant survival on the platform over a longer period of time.

We use a linear probability model of the following form to estimate the relationship between the demand shock experienced during the immediate period following the imposition of the shelter-in-place orders and the likelihood of restaurant survival a year following the period of analysis:

$$Y_i = \beta_1 DemandShock_i + \beta_2 Controls_i + City\ FE_i + Cuisine\ FE_i + \varepsilon \qquad (7)$$

In this equation, *i* indexes the restaurant. *Y* is a binary variable that is set equal to one if the restaurant is still active on the Uber Eats platform one year following the end of our period of analysis. We define a restaurant as active on the platform if they have offered supply on the Uber Eats platform within 30 or 90 days from the May 1, 2021, cut-off date. Granted, this is a crude measure of restaurant survival, as it primarily reflects activity on the platform rather than capturing the overall activity of the restaurant a year later. Nevertheless, we believe it is an important first step to probe whether the size of the increase in platform activity during this period of crisis is correlated with whether the restaurant is still active at all on the platform a year after the period of analysis.

Our main independent variable *Demand Shock* measures the size of the post-COVID demand shock experienced by sample restaurants, calculated as the count of orders received in the post-period divided by count of orders received in the pre-period. Thus, the analysis only includes restaurants that received at least one order in the pre-period and may be understating the effects of platform activity during the crisis on post-lockdown survival since it does not consider restaurants that adopted Uber Eats because of the lockdown and survived as a consequence.

We include a number of controls for restaurant-level characteristics that may alternatively explain both its performance on the platform and its subsequent survival: the number of orders the restaurant received on the platform during the pre-shock period of the sample, the age of the restaurant at the start of the pre-period, and the average rating of the restaurant on the platform at the start of the pre-period. We also include city and cuisine fixed effects to control for idiosyncratic city- or cuisine-level effects.[12]

We present the results of this analysis in Table 6. We find a positive relationship between the platform-mediated demand shock and the likelihood of restaurant survival on the platform one year following the sample analysis period. The estimates in Column 1, considering a restaurant as active on the platform if they have offered supply on the Uber Eats platform within 30 days of May 1, 2021, suggests that a restaurant that doubled its orders on the platform during the post-period (the lockdown period we study, March-May 2020) is 2.4 percentage points *more likely* to still be active on the platform more than a year following the enactment of the shelter-in-place orders due to the COVID pandemic than a restaurant did not experience an increase in orders on the platform during the post-period ($p = 0.01$). Note that a doubling of orders is not an unusual occurrence – recall, for instance, that the <u>average</u> San Francisco restaurant saw close to a 70% increase in demand during the post period.

When scaled by the unconditional sample mean of survival on the platform, this represents a 3.8% increase in the likelihood of survival on the platform. Results are consistent in statistical strength and magnitude if we consider a restaurant as still on the platform if they have offered supply on the Uber Eats platform within 90 days of May 1, 2021 (Column 2) and persist with ZIP code rather than city fixed effects (Columns 3 and 4).

Granted, interpreting the results of our analysis is subject to some caveats. The main dependent measure of restaurant survival is based only on activity on the platform, so thus this analysis does not capture restaurants that have survived off-platform even while cutting their activity on-platform. This is an especially important caveat, as restaurants that may not have been able to take advantage of the platform during this period of crisis may have had strategies that position themselves away from delivery or takeout models. Further, this analysis does not allow us to argue that a larger COVID-induced demand shock caused the increase in restaurant survival on the platform, as other unobservable restaurant characteristics may be playing a role.

[12] There are a number of caveats to this analysis, of course, which we discuss after presenting our results.

However, despite the caveats outlined above, we believe the findings suggest an important possible connection between the use of digital channels and small business resilience during crisis. Restaurants that were best able to leverage the platform to stimulate demand during the COVID-induced crisis were more likely to survive (at least on the platform) over a longer period of time. Crises that induce economic shocks seem increasingly frequent these days. Beyond the recent energy squeeze caused by the war in Ukraine, the future likely portents localized climate-related economic shocks that are more frequent. Our findings – that the ability to leverage a digital distribution channel can be an important source of resilience during periods of crisis – should form an important input as regulators consider the societal benefits and costs of these digital platforms when shaping regulatory policy.

## 7. CONCLUSION

As markets continue to rebound from the effects of the pandemic, "business-as-usual" has been transformed. There has been a significant and possibly permanent shift away from in-person commerce and towards digital interaction. This is not limited to the restaurant industry – apparel and accessories stores experienced a year-over-year decline of 89.3% in April 2020, which while understandable given shelter-in-place orders and pandemic-related fears, is staggering, perhaps the biggest YoY decline in any sector ever, and an inflection point that signals the future dominance of ecommerce in this sector (Tappe and Meyersohn, 2020).

In this paper, we document how local businesses were able to leverage an online delivery platform to stay alive and generate much-needed revenue during this period of crisis. We document that restaurants experienced an increase in demand on the platform following the imposition of shelter-in-place guidance, highlight how the shuttering of many restaurants affected market competition and shaped the size of the demand shock experienced, and explore heterogeneity in this effect based on restaurant and locality characteristics. In doing so, we uncover some of the mechanisms driving the demand shifts – for example, how newer and infrequent consumers played a key role, how the dynamics of the shift reveal early consumer caution about delivery followed by a more substantial increase in the later phases of the lockdown, and how the marginal benefits of the digital channel were greater for lower-income neighborhoods. Further, we present evidence that restaurants that experienced larger increases in demand on the platform during the shock were more likely to remain alive on the platform over a longer period of time, linking the size of the platform-mediated demand shock to longer-term success and survival and suggesting that the platform may have served as a source of resilience during this unprecedented crisis.

Summarizing our contributions: we provide clear evidence that digital platforms had a positive short-term effect on a specific category of small businesses. Our evidence is consistent with the use of these platforms being positively associated with survival post-crisis. We define a new measure of competition that groups competitors based on "revealed shared preference" of consumers, and which could find general applicability in future studies in lieu of more noisy groupings based on business category labels not explicitly derived from overlapping observed demand. We use this measure of competition to provide clear empirical evidence of an economic trade-off between market expansion and provider competitive intensity. Further, we discuss the strategic importance of managing these countervailing stakeholder incentives as a platform pursues rapid market expansion.

Like any study, ours has limitations that we have discussed when presenting our results. It would have been useful to have a control group to be able to get a better sense for the counterfactual outcomes, but in our analyses, all restaurants were exposed to the COVID-19 shock. Due to data limitations, we are unable to explore strategic changes that restaurants may have made to take advantage of the prevalence of this new channel or explore changes in consumer composition in detail. Further, our analysis is limited to five of the largest cities in the United States and does not include small towns or rural areas. Despite these limitations, our work sheds lights on important empirical patterns and provides a foundation that future work can build upon.

While the pandemic caused many restaurants to close their doors for good, those that survive will be the ones that were best able to leverage alternative distribution channels. As consumers become more accustomed to navigating digital delivery services and ordering meals online, a digital channel may be all but essential for restaurant survival. The connection between platform-sourced demand and survival will be indelible, likely leading the survivors to double down on digital, seeing it as a critical source of resilience.

All of these factors point to the heightened importance of platforms like Uber Eats in the economy of the future. Our findings provide insight into the details and dynamics of the role that such platforms play in mitigating the adverse effects of negative economic shocks, underscoring the risks associated with policy that may curtail their growth or reach while also shedding new light on the different economic factors at play when a business sells through a platform.

**Figure 1: Demand-Side Response to the COVID-19 Pandemic on the Uber Eats Platform**

*A. Consumer Sessions with Demand Intent.*

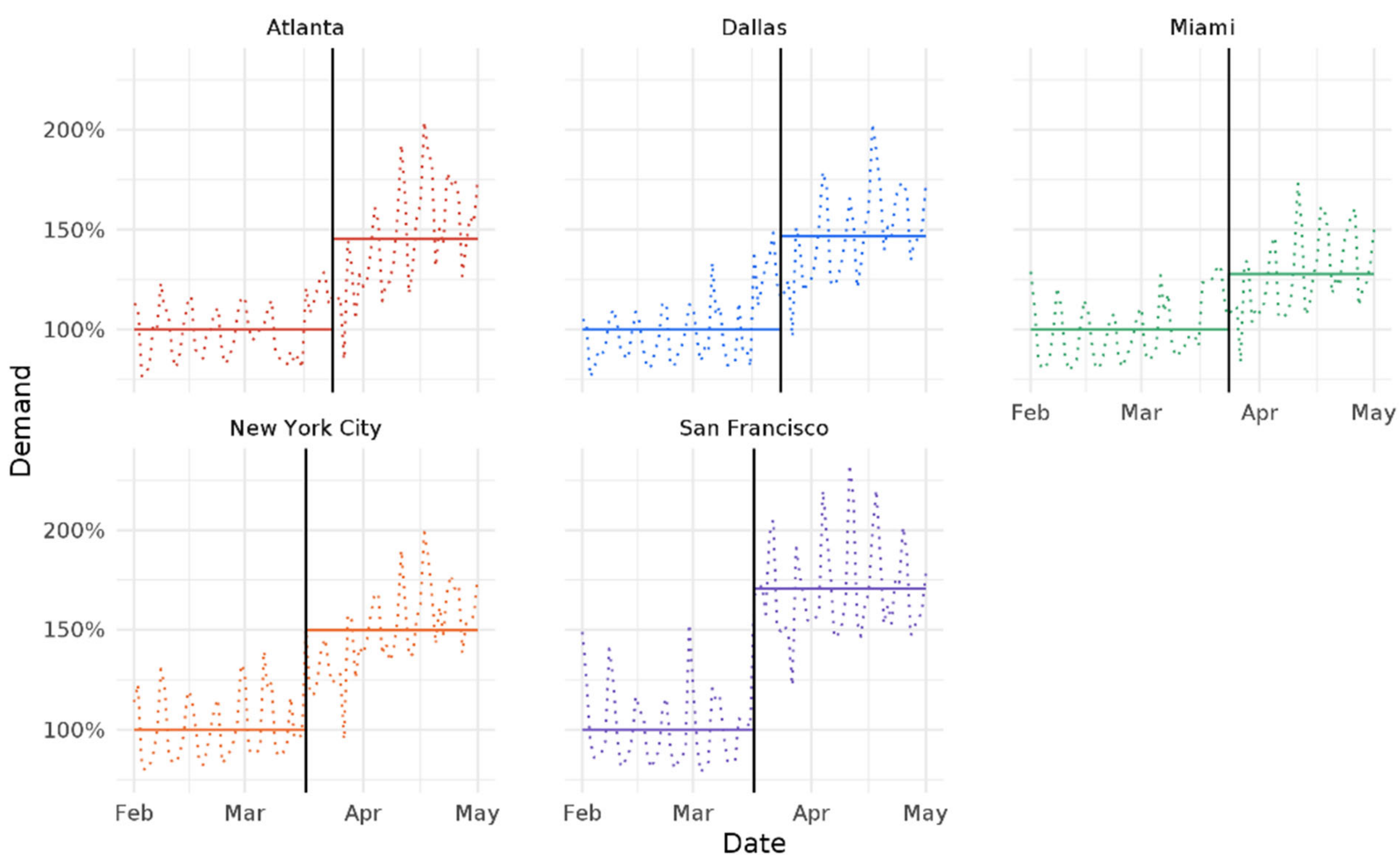

*B. Daily Completed Orders.*

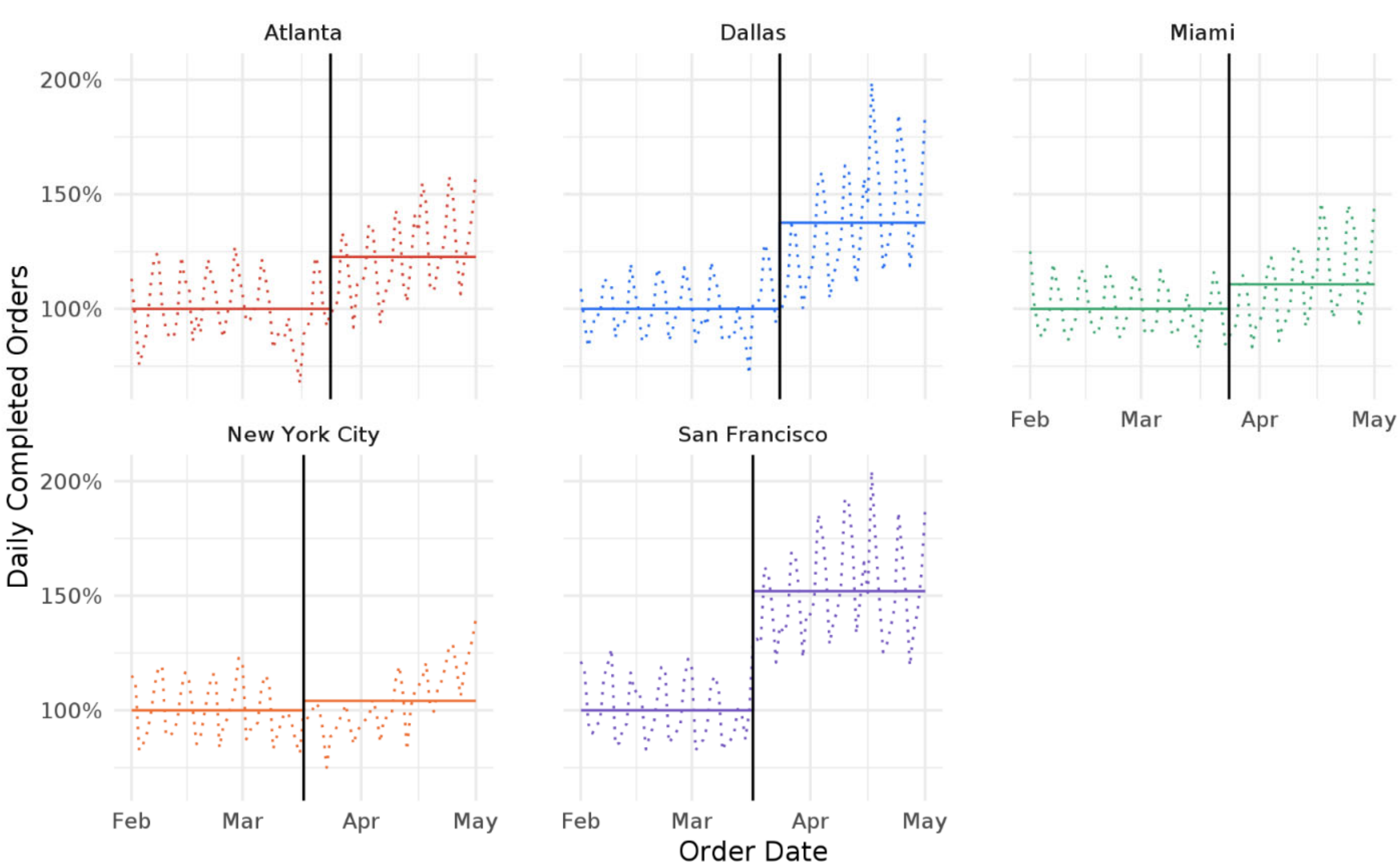

*C. Conversion Rate.*

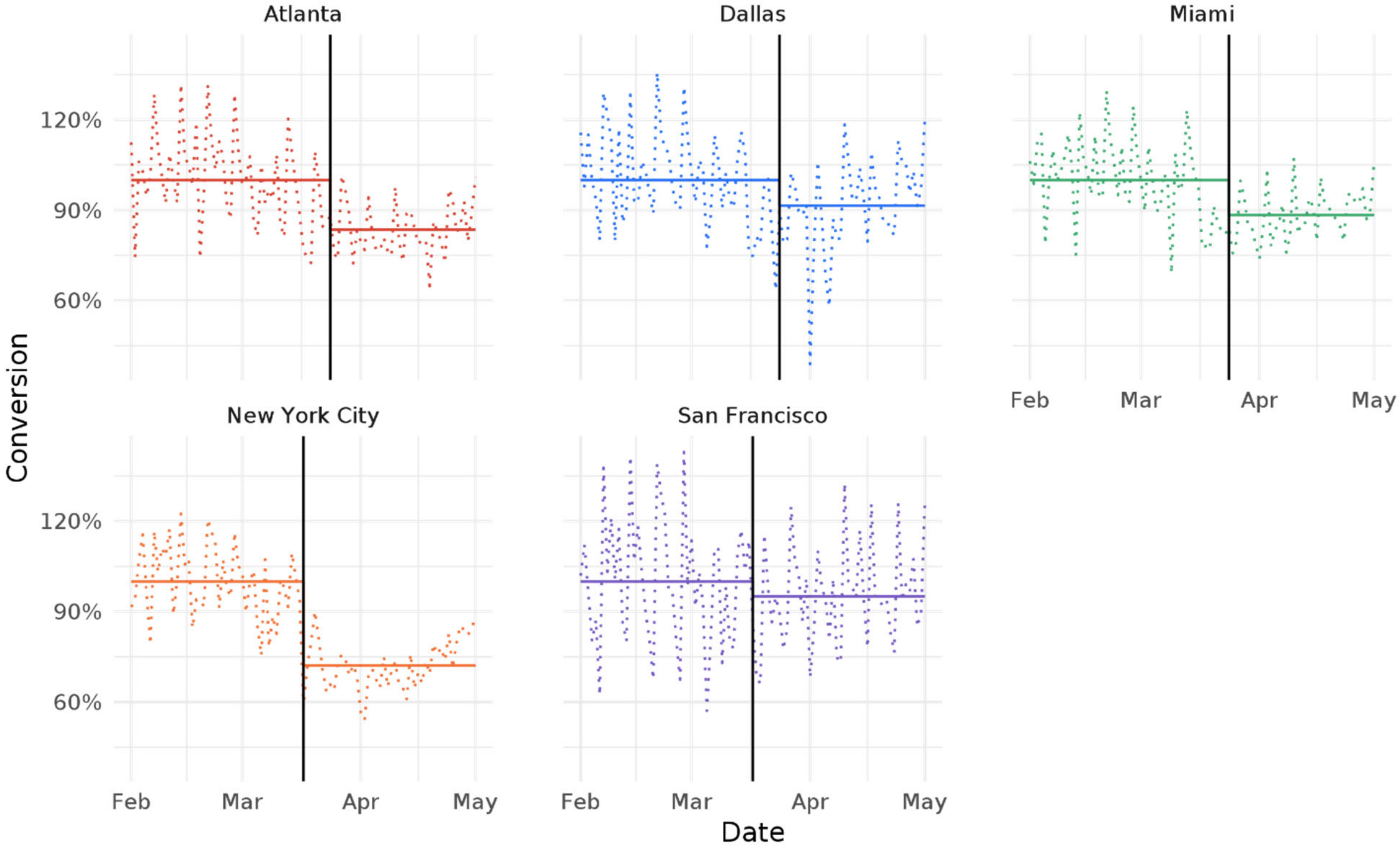

.

Note: In Panel A, the dotted curves plot the daily number of consumer sessions with demand intent between February 1 and May 1 across the five sample cities, normalized such that the average pre-lockdown value is equal to 1. In Panel B, the dotted curves plot the average daily number of completed orders in the pre-lockdown and post-lockdown time windows between February 1 and May 1 across the five sample cities, normalized such that the average pre-lockdown value is equal to 1. In Panel C, the dotted curves plot the average daily conversion rate in the pre-lockdown and post-lockdown time windows between February 1 and May 1 across the five sample cities, normalized such that the average pre-lockdown value is equal to 1. In all panels, the vertical line represents the day on which the shelter-in-place guidance was issued. The solid horizontal lines depict the average of each considered metric across the pre- and post-period.

**Figure 2: Supply-Side Response to COVID-19 Pandemic on the Uber Eats Platform**

*A. Number of Open Restaurants.*

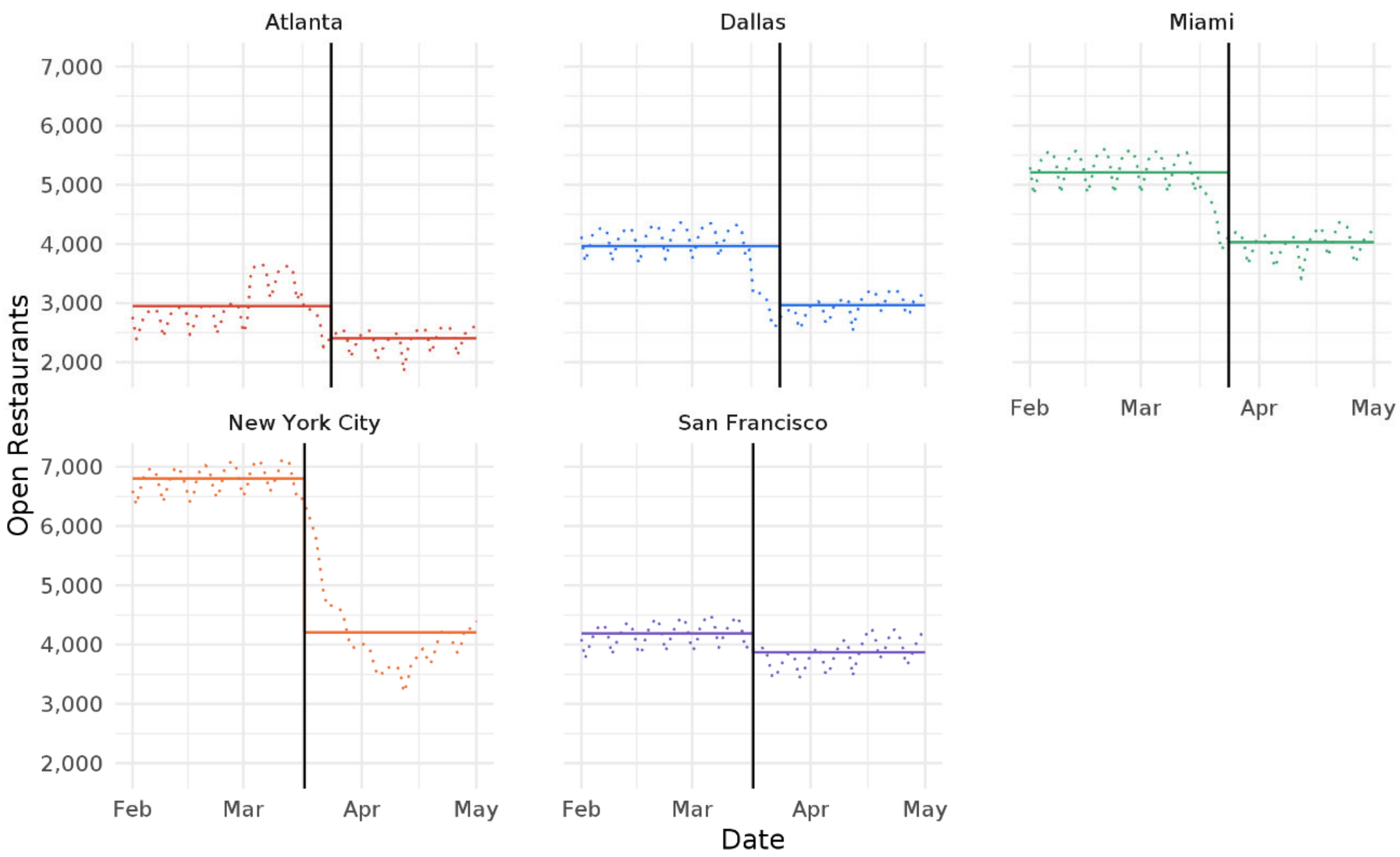

*B. Average Daily Supply Hours for Open Restaurants.*

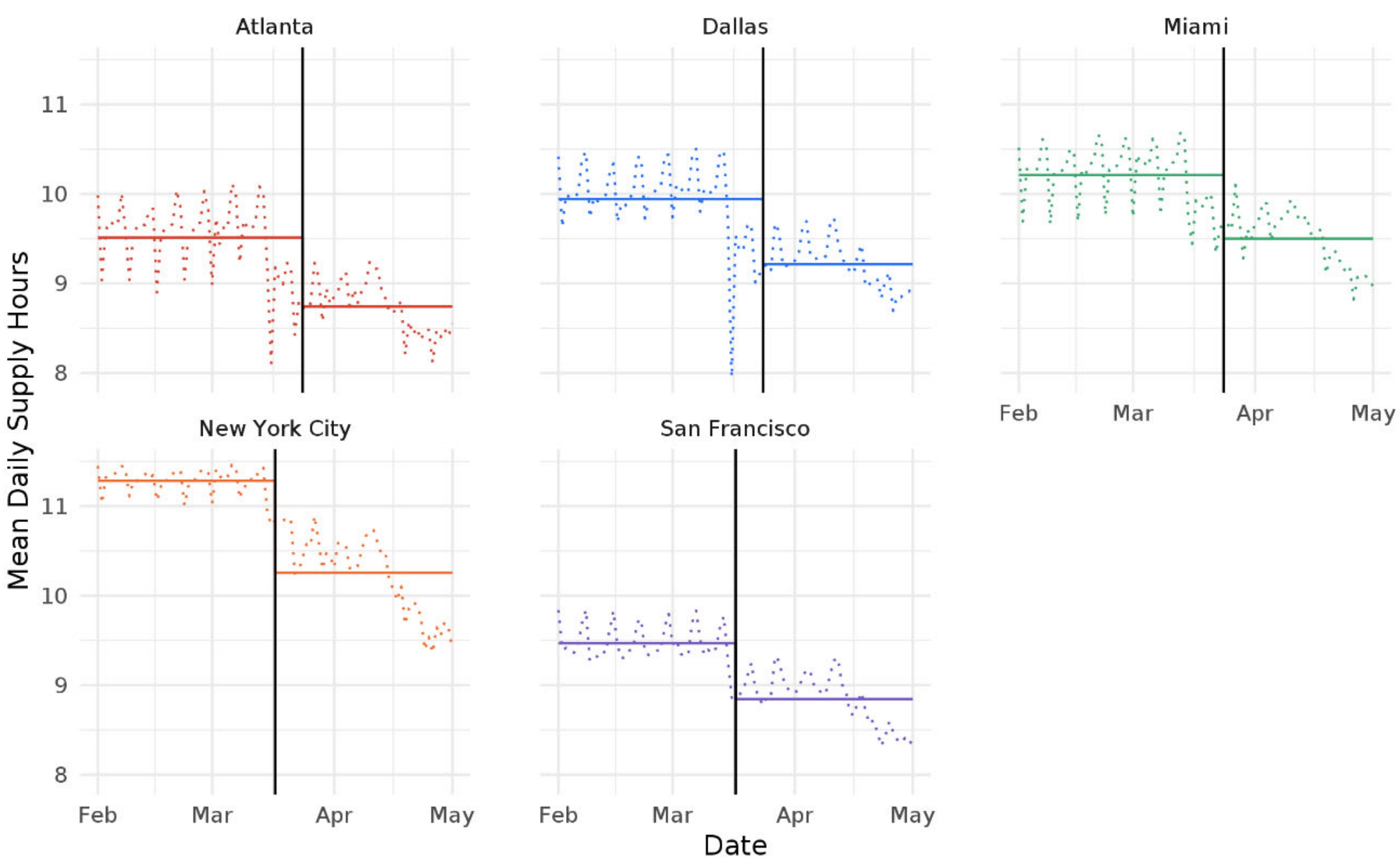

Note: In Panel A, the dotted curves plot the average number of restaurants that offer at least one hour of supply on the Uber Eats platform on any given day. In Panel B, the dotted curves plot the average number of daily supply hours per restaurant, only counting restaurants that offer supply on the Uber Eats platform on any given day. The vertical line represents the day on which the shelter-in-place guidance was issued. In both charts, the solid horizontal lines depict the average in the pre-lockdown and post-lockdown windows.

**Figure 3: Changes in Order Velocity on the Uber Eats Platform in Response to the COVID-19 Pandemic**

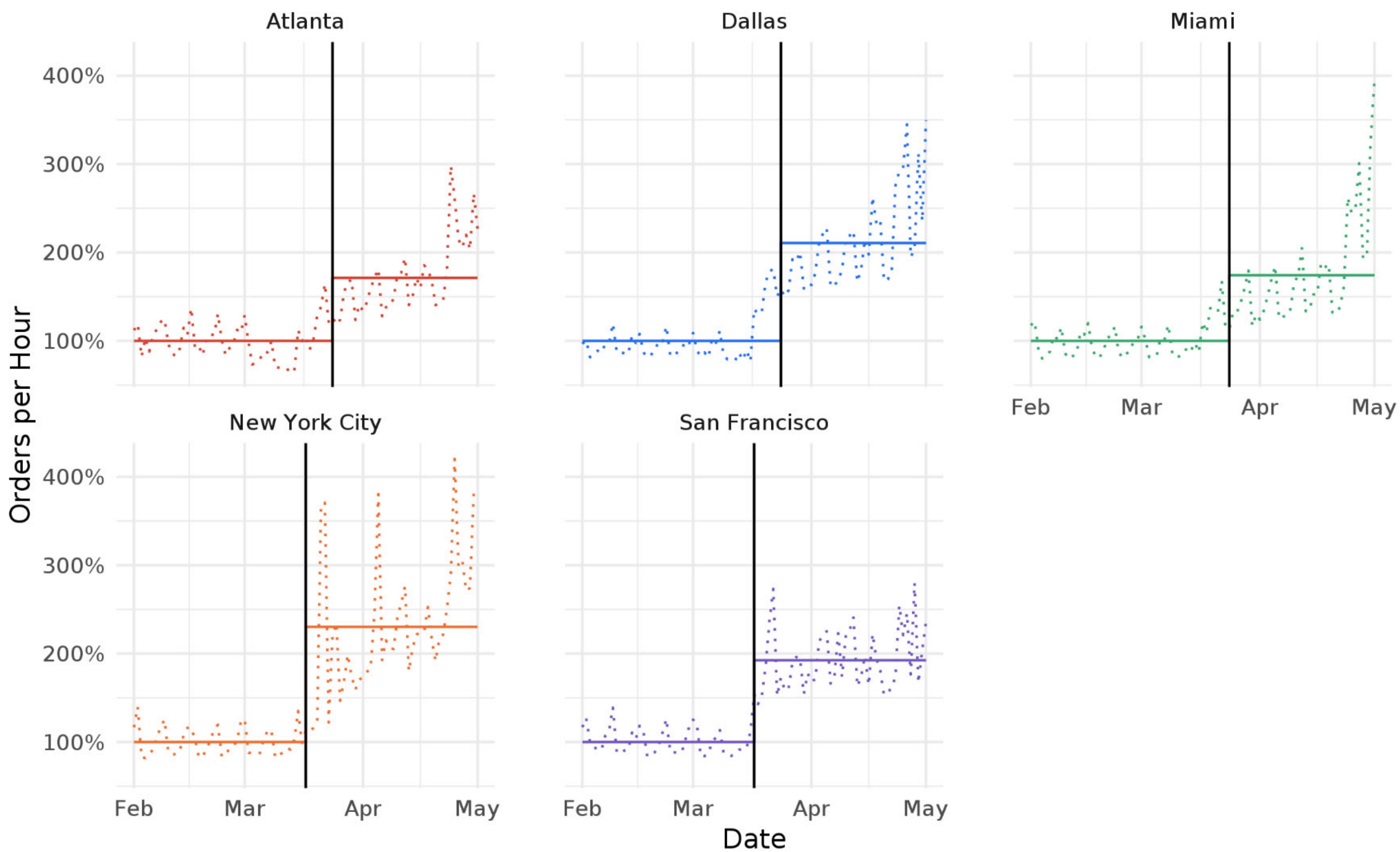

Note: The dotted curves plot the daily number of orders/supply hour (normalized velocity) fulfilled by restaurants between February 1 and May 1 across our five cities, normalized in a way that makes the average pre-lockdown value equal to 1. The vertical line represents the day on which the shelter-in-place guidance was issued. The solid horizontal lines depict the average normalized velocity in the pre-lockdown and post-lockdown time windows.

**Figure 4: Temporal Effect of Shelter-in-Place Guidance on Daily Orders on the Uber Eats Platform**

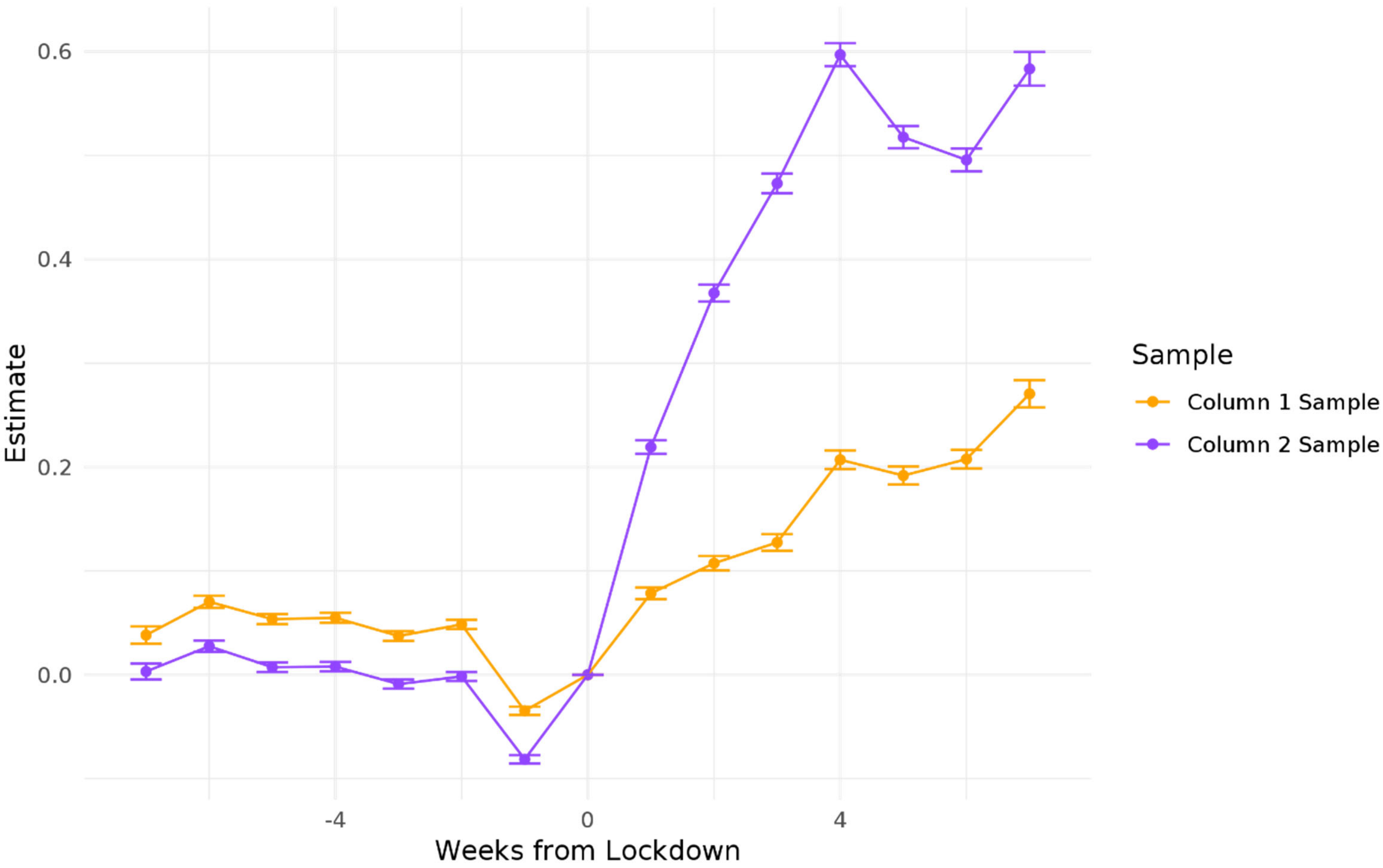

Note: This figure displays estimates of how the post-effect develops temporally in the weeks following the shelter-in-place orders. The dots represent point estimates, and the brackets are 95% confidence intervals. The y-intercept (weeks from lockdown set equal to zero) is the omitted week and represents the week prior to the shelter-in-place orders. The estimates underlying this figure are presented in *Appendix Table A7*.

**Figure 5: City-Specific Impact of Shelter-in-Place Guidance on Daily Orders on the Uber Eats Platform**

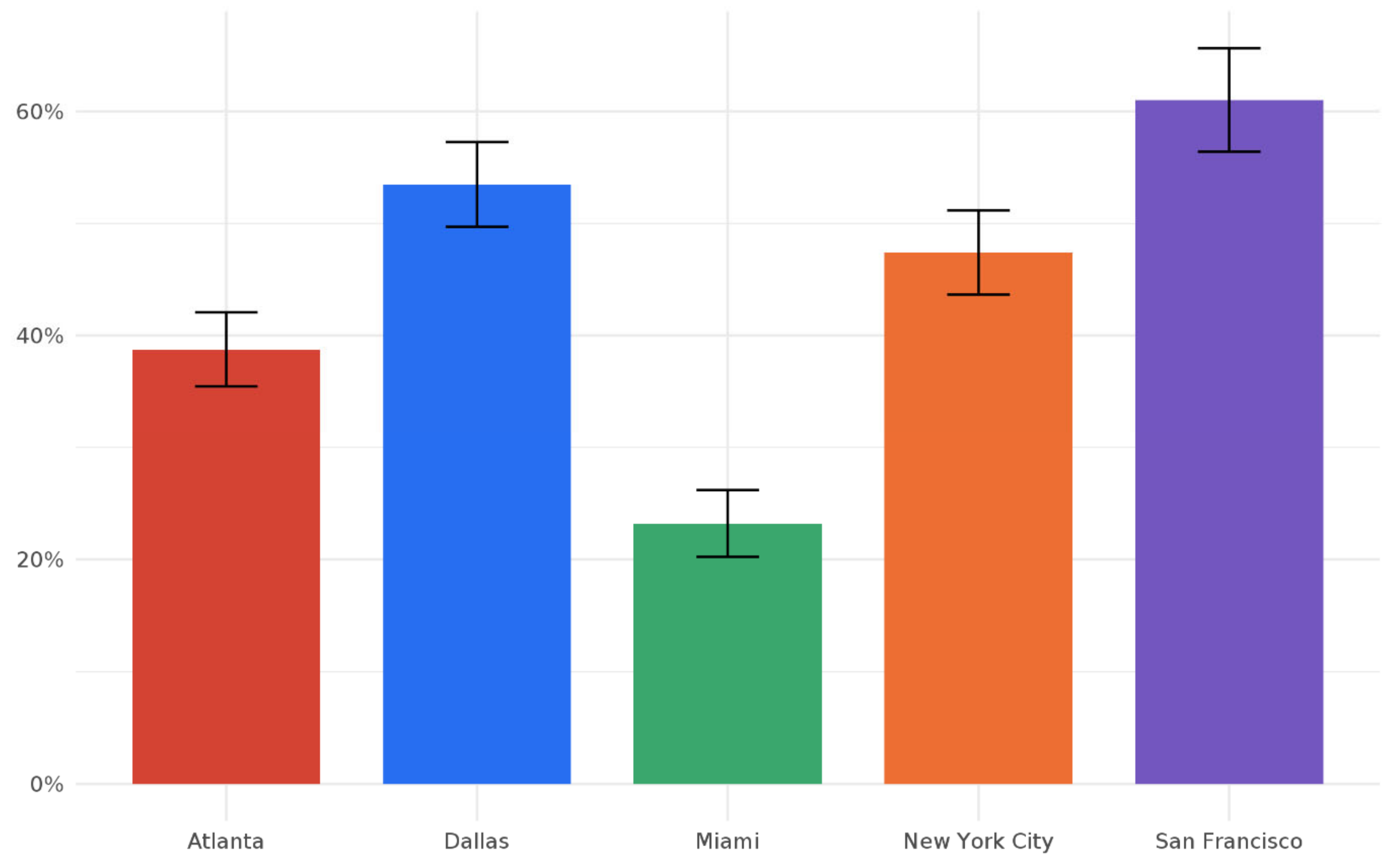

Note: The figure displays a bar chart displaying the city-specific marginal effect of shelter-in-place guidance on daily orders for restaurants offering supply on a given day. Each bar depicts, for the respective city, the percentage increase in average daily orders per restaurant in the post-lockdown period, relative to the average number of orders received by restaurants in that city during the pre-lockdown period. The error bars depict the 95% confidence interval. Estimates are derived from the regression results presented in Table 2.

**Figure 6: Post-Effect Across Different Levels of Competition.**

*A. Cuisine Measure.*

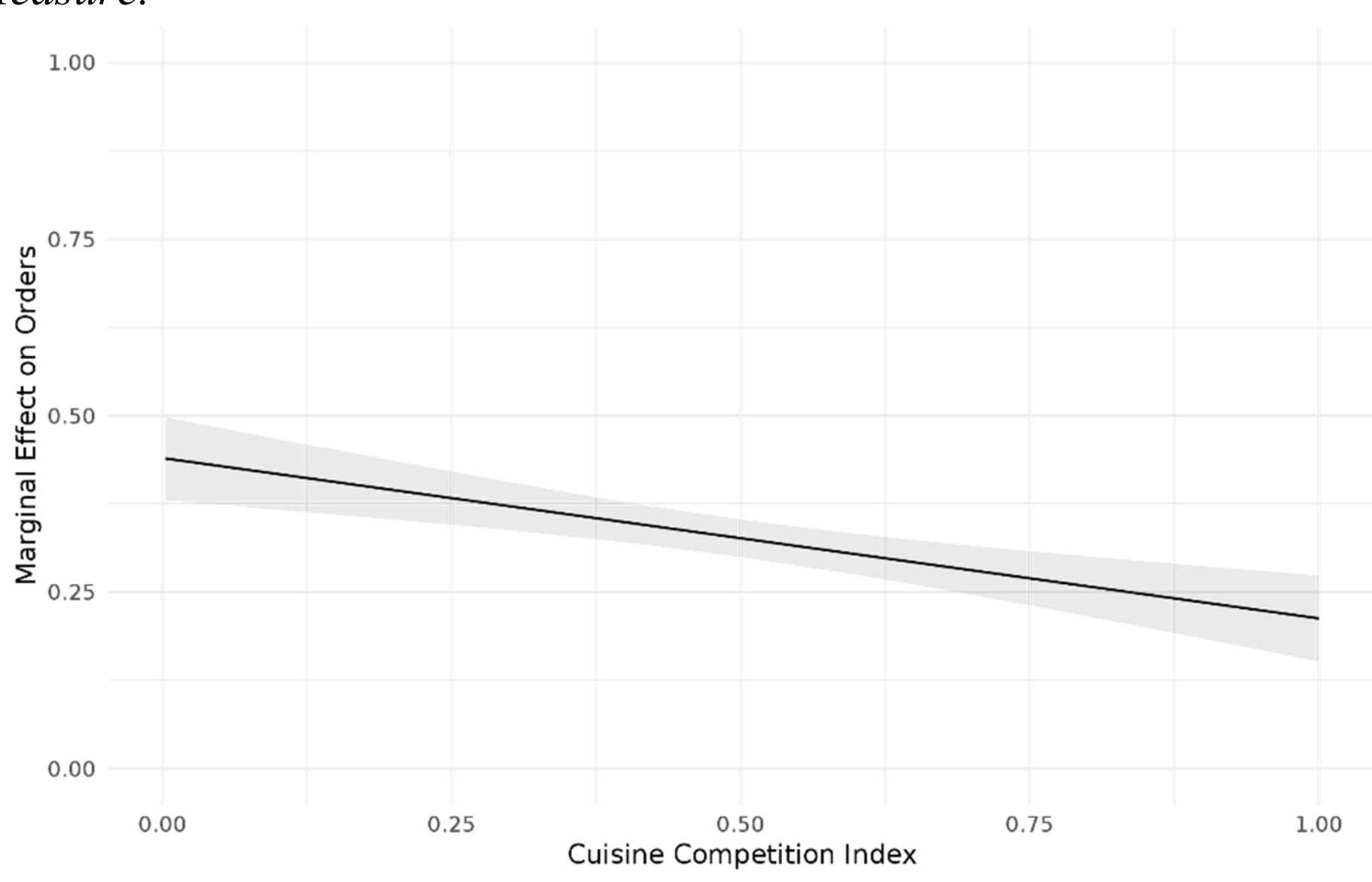

*B. Network Measure.*

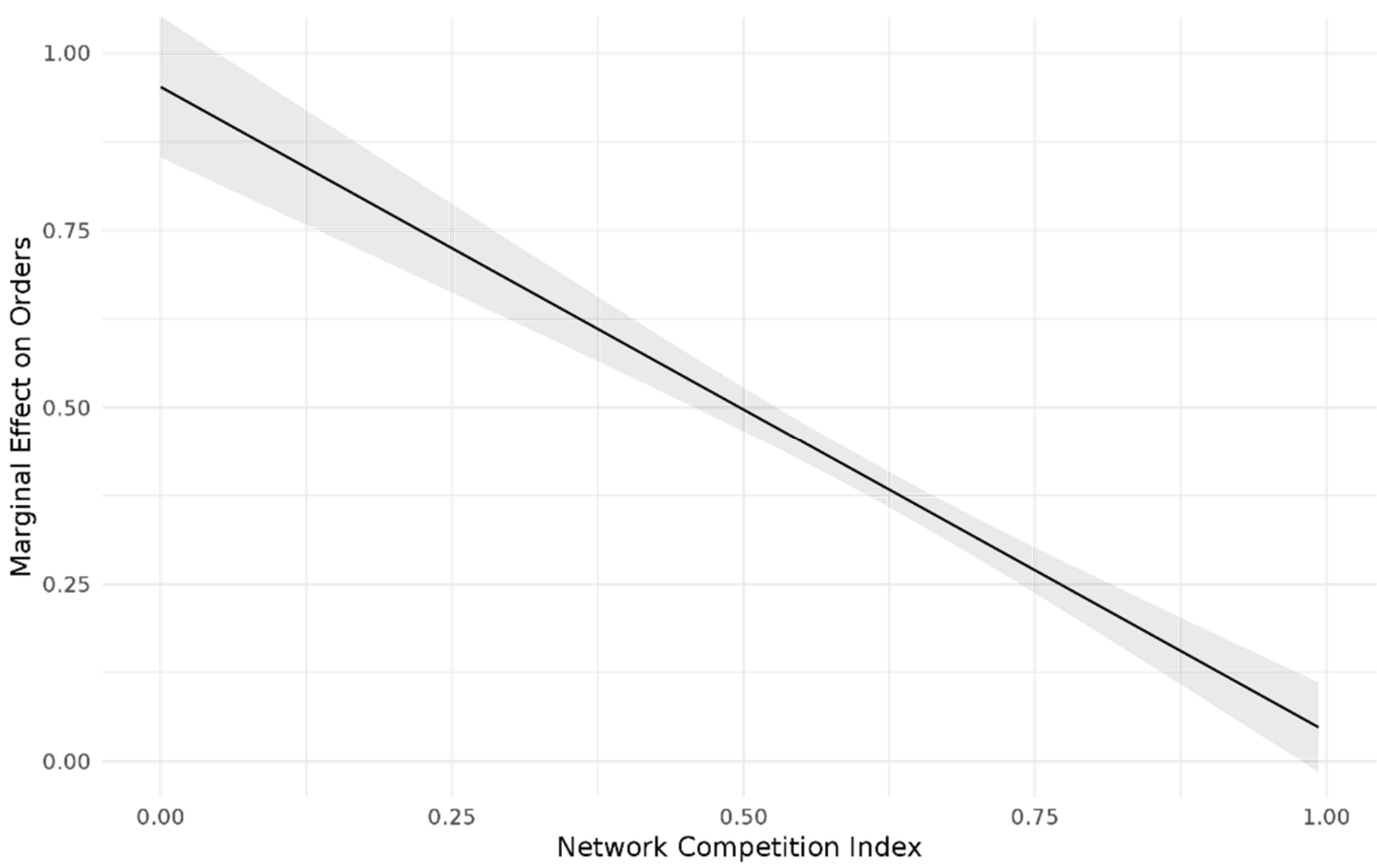

Note: The figure displays the estimate of the marginal effect of post-period on the count of restaurant orders at different levels of competition. Panel A measures competition using the cuisine competition index and Panel B measures competition using the network competition index. The estimates underlying these figures are presented in Table 4.

## Table 1: Summary Statistics

*Panel A. Restaurant Counts.*

| **Restaurant Counts** | |
|---|---|
| *On-Platform Pre- or Post-Period* | |
| Atlanta | 4,643 |
| Dallas | 5,604 |
| Miami | 7,163 |
| New York City | 9,089 |
| San Francisco | 6,264 |
| **Total** | **32,763** |
| | |
| *On-Platform Pre- and Post-Period* | |
| Atlanta | 3,016 |
| Dallas | 3,323 |
| Miami | 4,933 |
| New York City | 7,040 |
| San Francisco | 4,581 |
| **Total** | **22,893** |

*Panel B. Restaurant Supply on Uber Eats.*

| **Percent Open** | Whole Sample | Pre-Period | Post-Period |
|---|---|---|---|
| Atlanta | 58.5% | 63.5% | 51.8% |
| | *8.8%* | *8.1%* | *3.9%* |
| Dallas | 63.0% | 70.7% | 52.9% |
| | *10.9%* | *8.0%* | *3.4%* |
| Miami | 65.7% | 72.7% | 56.2% |
| | *9.4%* | *5.6%* | *3.1%* |
| New York City | 60.4% | 74.8% | 46.3% |
| | *15.5%* | *2.5%* | *8.0%* |
| San Francisco | 64.3% | 66.9% | 61.8% |
| | *4.4%* | *3.3%* | *3.8%* |
| **Total** | **62.4%** | **69.8%** | **53.8%** |

*Panel C. Orders on Uber Eats.*

| **Orders-Per-Day** | | | |
|---|---|---|---|
| *All Restaurant-Days* | | | |
| | Whole Sample | Pre-Period | Post-Period |
| Atlanta | 4.8 | 4.7 | 4.8 |
| | *9.3* | *8.8* | *9.8* |
| Dallas | 3.4 | 3.1 | 3.8 |
| | *7.3* | *6.6* | *8.0* |
| Miami | 5.5 | 5.4 | 5.5 |
| | *11.3* | *10.4* | *12.3* |
| New York City | 4.7 | 4.8 | 4.6 |
| | *12.3* | *10.8* | *13.6* |
| San Francisco | 3.5 | 3.0 | 4.0 |
| | *7.3* | *5.7* | *8.4* |
| **Total** | **4.5** | **4.3** | **4.6** |
| | | | |
| *Restaurant-Days with Supply* | | | |
| | Whole Sample | Pre-Period | Post-Period |
| Atlanta | 7.1 | 6.0 | 8.9 |
| | *10.6* | *9.5* | *12.0* |
| Dallas | 4.9 | 3.8 | 6.9 |
| | *8.3* | *7.1* | *9.7* |
| Miami | 7.7 | 6.7 | 9.5 |
| | *12.8* | *11.2* | *14.9* |
| New York City | 7.3 | 5.8 | 9.6 |
| | *14.7* | *11.6* | *18.3* |
| San Francisco | 4.8 | 3.7 | 6.0 |
| | *8.2* | *6.1* | *9.8* |
| **Total** | **6.5** | **5.3** | **8.2** |

*Panel D. Competition Measures.*

| **Competition Measures** | | | |
|---|---|---|---|
| *Cuisine Measure* | | | |
| | Whole Sample | Pre-Period | Post-Period |
| Atlanta | 54.1% | 60.7% | 43.2% |
| | *18.6%* | *17.4%* | *15.0%* |
| Dallas | 55.3% | 61.0% | 45.2% |
| | *17.5%* | *15.5%* | *16.4%* |
| Miami | 55.6% | 60.9% | 46.5% |
| | *18.2%* | *17.6%* | *15.3%* |
| New York City | 52.9% | 61.0% | 39.9% |
| | *19.4%* | *16.7%* | *15.9%* |
| San Francisco | 55.2% | 60.3% | 49.8% |
| | *15.5%* | *15.0%* | *14.1%* |
| **Total** | **54.5%** | **60.8%** | **44.9%** |
| *Network Measure* | | | |
| | Whole Sample | Pre-Period | Post-Period |
| Atlanta | 63.4% | 67.9% | 55.4% |
| | *11.0%* | *9.7%* | *8.4%* |
| Dallas | 62.6% | 66.3% | 54.9% |
| | *10.6%* | *9.7%* | *7.8%* |
| Miami | 70.0% | 73.7% | 62.8% |
| | *9.7%* | *8.5%* | *7.5%* |
| New York City | 63.1% | 71.3% | 48.9% |
| | *14.1%* | *7.9%* | *10.9%* |
| San Francisco | 61.9% | 66.5% | 56.3% |
| | *11.3%* | *10.7%* | *9.4%* |
| **Total** | **64.4%** | **69.7%** | **55.4%** |

*Panel E. Moderators.*

| | Average price | Average rating | Uber age | Mean household income | Smartphone penetration | Broadband penetration | Population Density |
|---|---|---|---|---|---|---|---|
| Atlanta | $23 | 4.43 | 1.24 | $75,000 | 88% | 87% | 2,745 |
| | *$7* | *0.39* | *1.21* | *$24,834* | *5%* | *8%* | *1,773* |
| Dallas | $22 | 4.49 | 1.11 | $79,988 | 88% | 88% | 3,506 |
| | *$8* | *0.43* | *1.08* | *$33,264* | *7%* | *9%* | *1,984* |
| Miami | $26 | 4.47 | 1.37 | $62,003 | 84% | 82% | 6,937 |
| | *$10* | *0.45* | *1.12* | *$20,957* | *6%* | *10%* | *5,932* |
| New York City | $23 | 4.39 | 1.43 | $83,069 | 82% | 83% | 53,824 |
| | *$9* | *0.45* | *1.10* | *$35,587* | *7%* | *9%* | *30,474* |
| San Francisco | $25 | 4.47 | 1.40 | $112,289 | 87% | 89% | 11,616 |
| | *$8* | *0.40* | *1.23* | *$38,024* | *4%* | *6%* | *11,842* |
| **Total** | **$24** | **4.44** | **1.33** | **$82,485** | **85%** | **85%** | **19,765** |

Note: The table presents summary statistics for variables used in the analysis. Panel A presents the count of unique restaurants, Panel B presents information regarding the percent of restaurants offering supply on the Uber Eats platform (defined as restaurants offering supply on a given day divided by restaurants that were open during the pre- or post-period in the sample city), Panel C presents information regarding average daily orders by restaurants on the Uber Eats platform, Panel D presents information regarding the competition measures used in the analysis, and Panel D presents information regarding the moderator variables used in the analysis. In Panels B-E, standard deviations are presented below means.

**Table 2: Regression Estimate of the Change in Daily Restaurant Orders on the Uber Eats Platform in Response to the COVID-19 Pandemic**

| | (1) | (2) | (3) |
|---|---|---|---|
| Sample: | All restaurant-days | Restaurant-days with supply | Restaurant-days with supply |
| Dependent variable: | Daily orders | Daily orders | Daily orders |
| Post | 0.1252*** | 0.4458*** | 0.3877*** |
| | *(0.0078)* | *(0.0089)* | *(0.0169)* |
| Post x Dallas | | | 0.1473*** |
| | | | *(0.0256)* |
| Post x Miami | | | -0.1556*** |
| | | | *(0.0227)* |
| Post x New York City | | | 0.0865*** |
| | | | *(0.0255)* |
| Post x San Francisco | | | 0.2226*** |
| | | | *(0.0290)* |
| | | | |
| Restaurant fixed effects? | Yes | Yes | Yes |
| Day of week fixed effects? | Yes | Yes | Yes |
| | | | |
| R-squared | 73.9% | 79.4% | 79.5% |
| Restaurant-day observations | 2,727,709 | 1,862,748 | 1,862,748 |

Note: The table presents regression results that assess the impact of shutdown orders on daily order counts on the Uber Eats platform. The variable *Post* takes the value 0 for all observations February 1 through the declaration of shelter-in-place guidance in a given city, and the value 1 for all observations following the declaration of shelter-in-place guidance through May 1. The data set in Column 1 is a balanced panel consisting of all restaurants that offered supply on at least one day between February 1 through May 1. The data set in Column 2 is an unbalanced panel consisting of only those restaurant-days for which the restaurant offered positive supply on that day. In column 3, we use the same sample as Column 2 but interact the *Post* variable with an indicator variable for each city. The dependent variable is scaled by the within-city pre-period mean. Standard errors are clustered at the restaurant-level. ***, **, and * represents significance at $p <$ 0.001, 0.01, and 0.05, respectively.

**Table 3: The Effect of Competition on Daily Restaurant Orders Pre- and Post- Shelter-in-Place Guidance**

| | (1) | (2) | (3) | (4) |
|---|---|---|---|---|
| Sample: | Restaurant-days with supply | Restaurant-days with supply | Restaurant-days with supply | Restaurant-days with supply |
| Dependent variable: | Daily orders | Daily orders | Daily orders | Daily orders |
| Post | 0.3200*** | 0.4389*** | 0.3725*** | 0.9521*** |
| | *-0.0137* | *-0.0305* | *(0.0128)* | *(0.0507)* |
| Cuisine measure of competition | -0.7306*** | -0.6336*** | | |
| | *(0.0755)* | *(0.0752)* | | |
| Post x Cuisine measure of competition | | -0.2265*** | | |
| | | *(0.0552)* | | |
| Network measure of competition | | | -0.5582*** | -0.0134 |
| | | | *(0.0811)* | *(0.0835)* |
| Post x Network measure of competition | | | | -0.9104*** |
| | | | | *(0.0793)* |
| Restaurant fixed effects? | Yes | Yes | Yes | Yes |
| Day of week fixed effects? | Yes | Yes | Yes | Yes |
| R-squared | 79.4% | 79.4% | 79.6% | 79.6% |
| Restaurant-day observations | 1,862,748 | 1,862,748 | 1,671,821 | 1,671,821 |

Note: The table presents regression results that assess the variation across cities in the impact of shutdown orders on daily order counts on the Uber Eats platform. This analysis is analogous to Table 1, except the *Post* variable is interacted with constructed measures of competition. The data set is an unbalanced panel consisting of only those restaurant-days for which the restaurant offered positive supply on that day. The dependent variable is scaled by the within-city pre-period mean. Standard errors are clustered at the restaurant-level. ***, **, and * represents significance at $p < 0.001$, 0.01, and 0.05, respectively.

**Table 4: Regression Estimate of the Effect of Supply on Total Orders in a Locality**

| | (1) | (2) | (3) | (4) | (5) | (6) | (7) | (8) | (9) |
|---|---|---|---|---|---|---|---|---|---|
| Sample: | All sample | All sample | All sample | Pre-period | Pre-period | Pre-period | Post-period | Post-period | Post-period |
| Dependent variable: | Log daily orders | Log daily orders | Log daily orders | Log daily orders | Log daily orders | Log daily orders | Log daily orders | Log daily orders | Log daily orders |
| Restaurants open | 0.0167*** | | | 0.0164*** | | | 0.0237*** | | |
| | *(0.0017)* | | | *(0.0028)* | | | *(0.0053)* | | |
| Log restaurants open | | 0.6842*** | | | 0.5714*** | | | 0.7766*** | |
| | | *(0.0620)* | | | *(0.0834)* | | | *(0.1021)* | |
| Percent restaurants open | | | 1.159*** | | | 0.8525*** | | | 1.353*** |
| | | | *(0.0691)* | | | *(0.0809)* | | | *(0.1306)* |
| ZIP code fixed effects? | Yes | Yes | Yes | Yes | Yes | Yes | Yes | Yes | Yes |
| Date fixed effects? | Yes | Yes | Yes | Yes | Yes | Yes | Yes | Yes | Yes |
| R-squared | 94.8% | 95.2% | 95.1% | 96.5% | 96.6% | 96.5% | 95.3% | 95.8% | 95.6% |
| ZIP code-day observations | 82,743 | 82,743 | 82,743 | 44,362 | 44,362 | 44,362 | 38,381 | 38,381 | 38,381 |

Note: The table presents regression results that assess the relationship between restaurant supply on the Uber Eats platform and total demand on the platform. The data set is a balanced panel consisting of ZIP code-days. The dependent variable is the log-transformed count of orders on the Uber Eats platform to restaurants located within the ZIP code. Standard errors are clustered at the ZIP code-level. ***, **, and * represents significance at $p < 0.001$, 0.01, and 0.05, respectively.

**Table 5: Heterogeneity in the Change in Daily Restaurant Orders on the Uber Eats Platform in Response to the COVID-19 Pandemic**

| | (1) | (2) | (3) | (4) | (1) | (3) | (4) | (4) |
|---|---|---|---|---|---|---|---|---|
| Sample: | Restaurant-days with supply | Restaurant-days with supply | Restaurant-days with supply | Restaurant-days with supply | Restaurant-days with supply | Restaurant-days with supply | Restaurant-days with supply | Restaurant-days with supply |
| Dependent variable: | Daily orders | Daily orders | Daily orders | Daily orders | Daily orders | Daily orders | Daily orders | Daily orders |
| Post | 0.4458*** | 0.4526*** | -0.2649*** | 0.4084*** | 1.481*** | 0.7458*** | 0.4388*** | 0.2197*** |
| | *(0.0089)* | *(0.0261)* | *(0.0714)* | *(0.0150)* | *(0.2464)* | *(0.1172)* | *(0.0950)* | *(0.0490)* |
| Post x Average meal price | | -0.0002 | | | | | | |
| | | *(0.0010)* | | | | | | |
| Post x Average rating | | | 0.1607*** | | | | | |
| | | | *(0.0164)* | | | | | |
| Post x Uber age | | | | 0.0232** | | | | |
| | | | | *(0.0084)* | | | | |
| Post x Log mean household income | | | | | -0.0894*** | | | |
| | | | | | *(0.0213)* | | | |
| Post x Smartphone penetration | | | | | | -0.3500** | | |
| | | | | | | *(0.1357)* | | |
| Post x Broadband penetration | | | | | | | 0.0097 | |
| | | | | | | | *(0.1089)* | |
| Post x Log population density | | | | | | | | 0.0254*** |
| | | | | | | | | *(0.0058)* |
| Restaurant fixed effects? | Yes | Yes | Yes | Yes | Yes | Yes | Yes | Yes |
| Day of week fixed effects? | Yes | Yes | Yes | Yes | Yes | Yes | Yes | Yes |
| R-squared | 79.4% | 79.6% | 79.5% | 79.3% | 79.4% | 79.4% | 79.4% | 79.4% |
| Restaurant-day observations | 1,862,748 | 1,679,072 | 1,640,239 | 1,851,010 | 1,844,598 | 1,845,829 | 1,845,829 | 1,845,560 |

Note: The table presents regression results that assess the variation across cities in the impact of shutdown orders on daily order counts on the Uber Eats platform. This analysis is analogous to Table 1, except the *Post* variable is interacted with various moderator variables. The data set is an unbalanced panel consisting of only those restaurant-days for which the restaurant offered positive supply on that day. The dependent variable is scaled by the within-city pre-period mean. Standard errors are clustered at the restaurant-level. ***, **, and * represents significance at $p < 0.001$, 0.01, and 0.05, respectively.

**Table 6: Demand-Side Shock and Long-Term Survival on the Platform**

| | (1) | (2) | (3) | (4) |
|---|---|---|---|---|
| Dependent variable: | Survived on platform (30-day definition) | Survived on platform (90-day definition) | Survived on platform (30-day definition) | Survived on platform (90-day definition) |
| Demand-shock | 0.0239* | 0.0239** | 0.0234* | 0.0233** |
| | *(0.0052)* | *(0.0044)* | *(0.0058)* | *(0.0050)* |
| Number of orders pre-shock | 0.0635*** | 0.0588*** | 0.0644** | 0.0594** |
| | *(0.0067)* | *(0.0065)* | *(0.0077)* | *(0.0073)* |
| Uber age | 0.0355* | 0.0307* | 0.0386* | 0.0333* |
| | *(0.0097)* | *(0.0081)* | *(0.0095)* | *(0.0078)* |
| Average rating | 0.0804** | 0.0751** | 0.0730** | 0.0691** |
| | *(0.0096)* | *(0.0093)* | *(0.0097)* | *(0.0108)* |
| City fixed effects? | Yes | Yes | No | No |
| Zip code fixed effects? | No | No | Yes | Yes |
| Cuisine fixed effects? | Yes | Yes | Yes | Yes |
| Price category fixed effects? | Yes | Yes | Yes | Yes |
| R-squared | 10.7% | 9.6% | 16.1% | 14.9% |
| Restaurant-day observations | 23,523 | 23,523 | 23,120 | 23,120 |

Note: The table presents regression results that assess how the size of the demand shock experienced on the Uber Eats platform following the shelter-in-place guidance is related to the likelihood that the restaurant remained on the Uber Eats platform a year following the sample period. The independent variable measures the size of the demand shock experienced following the shelter-in-place order, defined as the count of orders in the post-period divided by the pre-period. The dependent variable is a binary variable that takes the value of 1 if the restaurant has offered supply on the Uber Eats platform within 30 or 90 days of May 1, 2021. Standard errors are clustered at the city-level. ***, **, and * represents significance at $p < 0.001$, 0.01, and 0.05, respectively.

# Appendix for COVID-19 and Digital Resilience: Evidence from Uber Eats

**Appendix Figure A1: Year-over-Year Percent Change in Seated Diners in Response to the COVID-19 Pandemic**

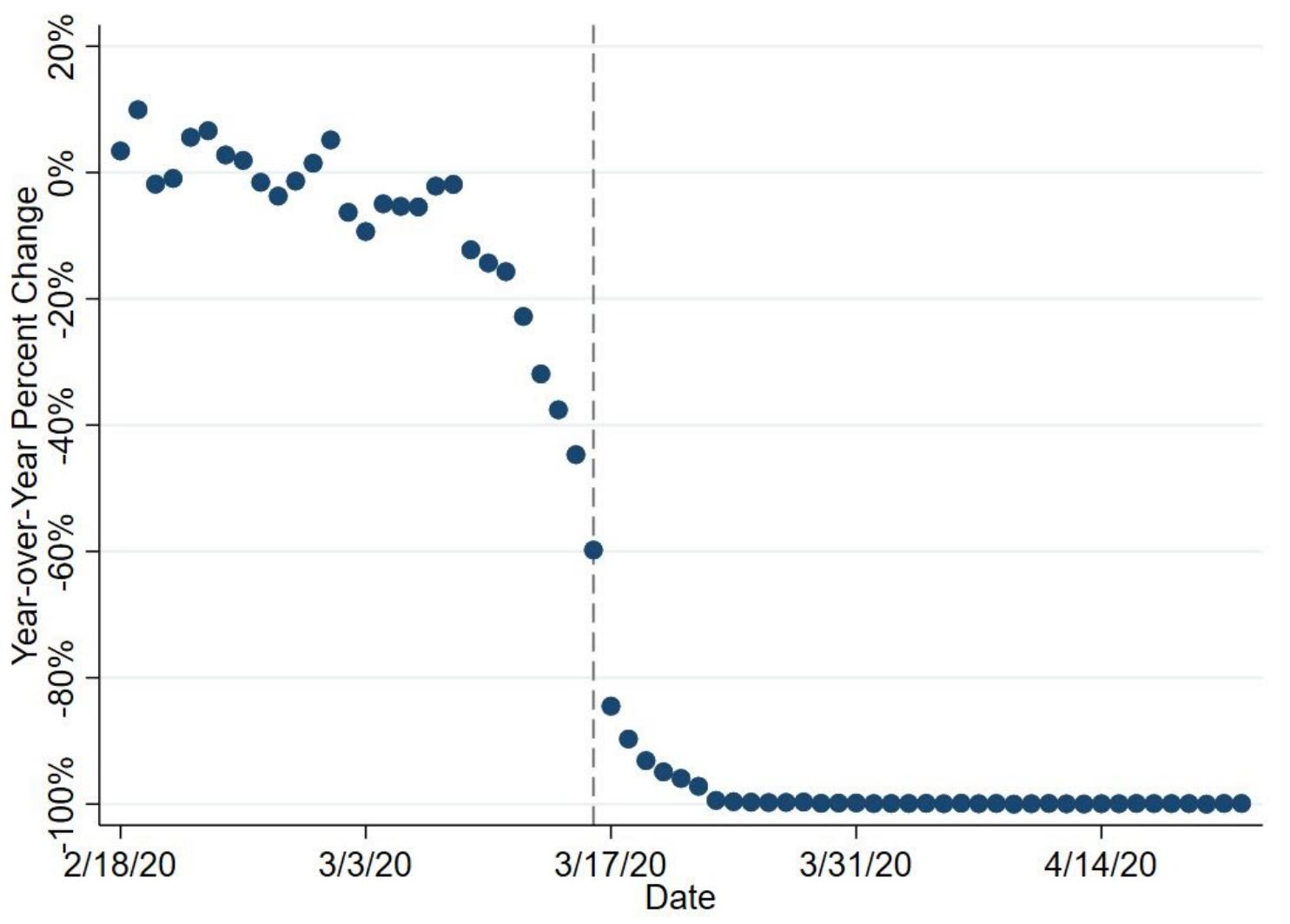

Note: Illustrates the year-over-year percent change in seated diners at restaurants on the OpenTable network in the United States The dashed line represents March 16.

## Appendix Figure A2: Distribution of Competition Index

*A. Whole Sample.*

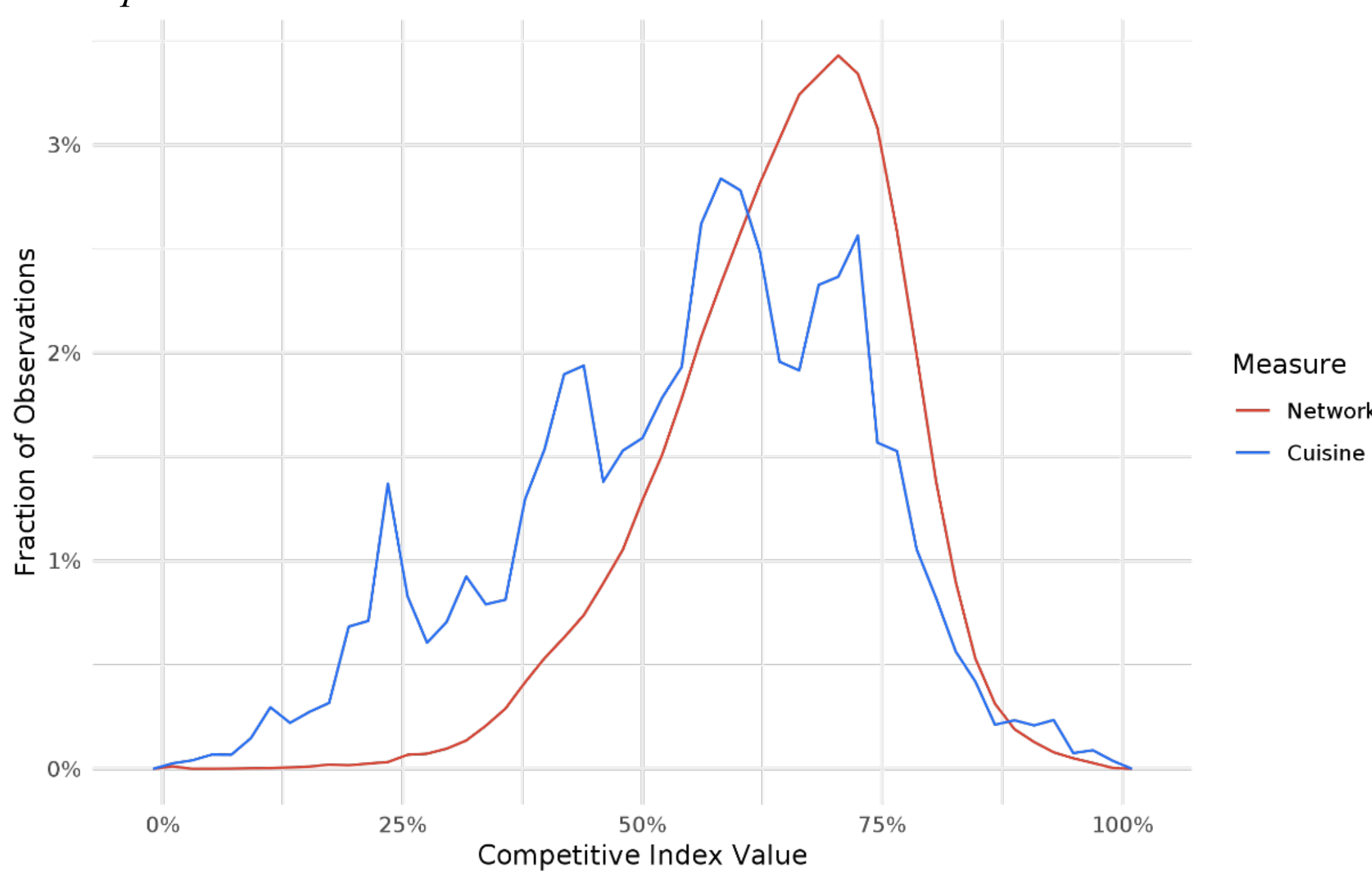

*B. Pre-Period.*

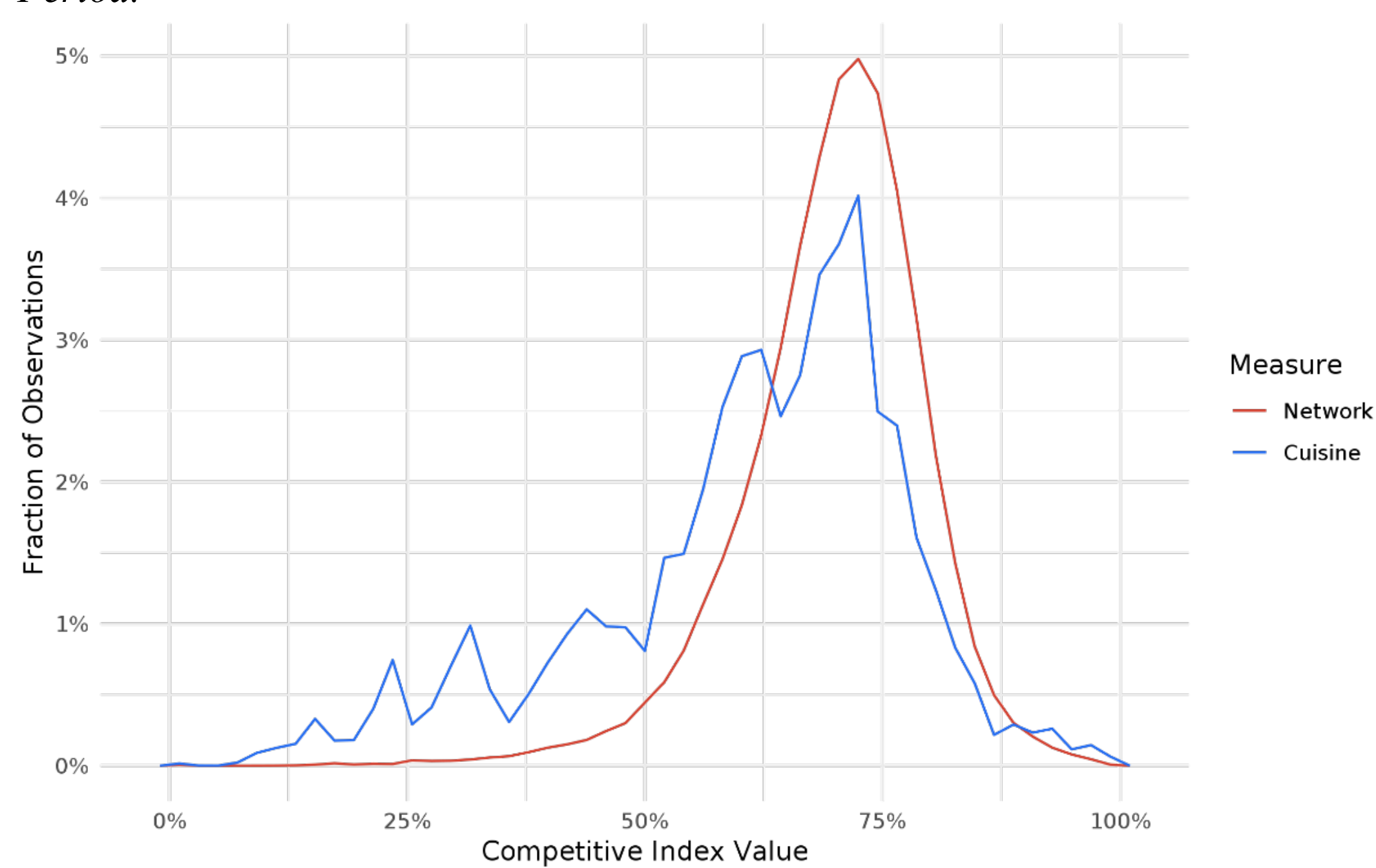

*C. Post-Period.*

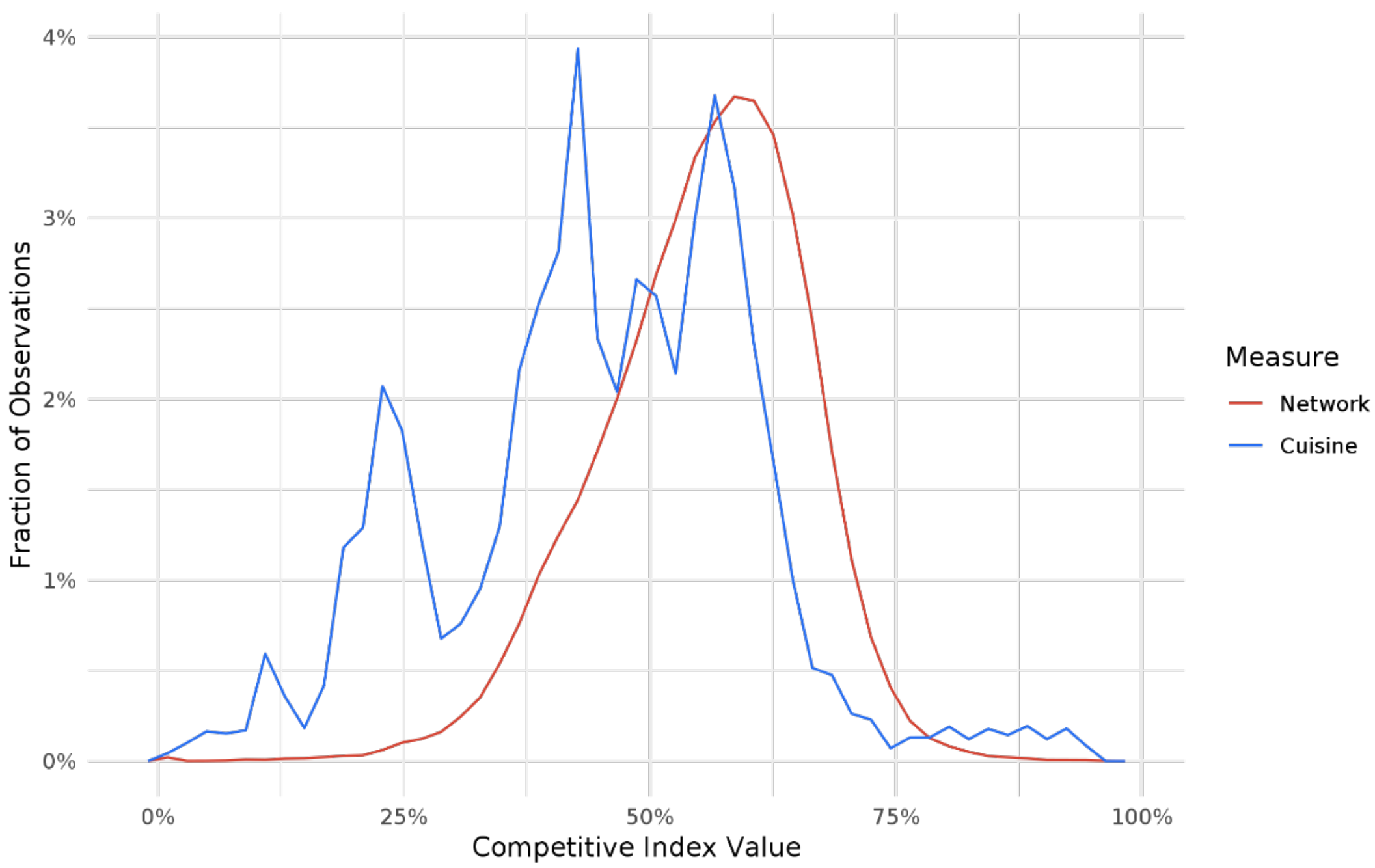

Note: The figure displays the distribution of the cuisine-level competition index. The competition index is measured as the percent of restaurants in a city within a cuisine that are open on a given day. This figure charts this distribution of observations at the city-cuisine-day level.

## Appendix Table A1: Variable Definitions

| Variable | Definition | Table |
|---|---|---|
| *Post* | An indicator variable that takes the value of 1 for all observations following the declaration of shelter-in-place guidance through May 1. See *Appendix Table A2* for city-specific timelines | Tables 2-3, 5 |
| *Daily orders (restaurant-level)* | The count of orders a restaurant receives in a day. For data anonymity purposes and to ease interpretation of coefficients, order counts at the restaurant-day level are scaled by the pre-period mean count of daily orders across all restaurants within the city. | Tables 2-3, 5 |
| *Cuisine measure of competition* | The percent of same-cuisine restaurants within a city offering supply on a given day. | Table 3 |
| *Network measure of competition* | The bipartite measure of competitive intensity measure described in the draft. The competitive intensity of restaurant *b* on restaurant *a* is the sum, across all customers of restaurant a, the likelihood each customer orders from restaurant *b* weighted by the importance of the customer to restaurant *a*. The measure takes the sum of the competitive intensity scores for each dyad the restaurant is in for which the competing restaurant is open on that day. | Table 3 |
| *Restaurants open* | The count of restaurants offering supply within the ZIP code on a given day. | Table 4 |
| *Log restaurants open* | The log-transformed count of restaurants offering supply within the ZIP code on a given day. | Table 4 |
| *Percent restaurants open* | The percent of restaurants within the ZIP code offering supply on a given day. | Table 4 |
| *Log daily orders (ZIP code-level)* | The log-transformed count of orders placed on the Uber Eats platform to restaurants located in the ZIP code. | Table 4 |
| *Average meal price* | The average price of a meal based on orders made to the restaurant on the Uber Eats platform at the start of the sample period. | Table 5 |
| *Average rating* | The average rating of the restaurant (out of 5) based on consumer evaluations on the Uber Eats platform at the start of the sample period. | Tables 5 and 6 |
| *Uber age* | The length of time that the restaurant has been on the Uber Eats platform from the start of the sample period. | Tables 5 and 6 |
| *Log mean household income* | The log mean household income at the ZIP-code level. | Table 5 |
| *Smartphone penetration* | The percent of the population at the ZIP-code level that reports owning a smartphone. | Table 5 |
| *Broadband penetration* | The percent of the population at the ZIP-code level that reports having a broadband internet connection at home. | Table 5 |
| *Log population density* | The log-transformed population density, calculated as population per square mile. | Table 5 |
| *Demand-shock* | The size of the demand shock experienced during the post-period, defined as total orders during the post-period divided by total orders during the pre-period. | Table 6 |
| *Number of orders pre-shock* | The number of orders that the restaurant received on the Uber Eats platform during the "pre" period, from February 1 to the imposition of shelter-in-place orders. | Table 6 |
| *Survived on platform* | An indicator variable that takes the value of 1 if the restaurant remains on the Uber Eats platform on May 1, 2021. We define a restaurant as remaining on the platform if it has offered supply within 30 or 90 days of May 1st. | Table 6 |

**Appendix Table A2: Pre- and Post-Lockdown Periods by Sample City**

| City | Pre-Lockdown Period | Post-Lockdown Period |
|---|---|---|
| Atlanta | February 1 – March 23 | March 24 – May 1 |
| Dallas | February 1 – March 23 | March 24 – May 1 |
| Miami | February 1 – March 23 | March 24 – May 1 |
| New York City | February 1 – March 16 | March 17 – May 1 |
| San Francisco | February 1 – March 16 | March 17 – May 1 |

Note: Date ranges are inclusive. Pre- and post-lockdown dates are defined by the date that the city announced "shelter-in-place" or "stay-at-home" guidance. Dates are sourced using the *New York Times* tracking information regarding dates at which states and cities have imposed stay-at-home directives (Mervosh et al. 2020). Dallas and Miami prohibited in-restaurant dining a week before instating shelter-in-place guidance. We use the shelter-in-place guidance for pre- and post-dates in these cities for consistency. Results are robust to using the date of restaurant closure as well.

**Appendix Table A3: Regression Estimate of the Change in Daily Restaurant Orders on the Uber Eats Platform in Response to the COVID-19 Pandemic with a Uniform Shelter-in-Place Date**

| | (1) | (2) |
|---|---|---|
| Sample: | All restaurant-days | Restaurant-days with supply |
| Dependent variable: | Daily orders | Daily orders |
| Post | 0.1149*** | 0.4122*** |
| | *(0.0078)* | *(0.0086)* |
| Restaurant fixed effects? | Yes | Yes |
| Day of week fixed effects? | Yes | Yes |
| R-squared | 73.9% | 79.3% |
| Restaurant-day observations | 2,727,709 | 1,862,748 |

Note: This analysis is analogous to Table 2 in the draft except that the indicator for the post-period is set equal to 1 for all sample cities on March 17. ***, **, and * represents significance at $p < 0.001$, 0.01, and 0.05, respectively.

## Appendix Table A4: Characteristics Predicting Restaurant Supply in the Post-Period

| | (1) | (2) |
|---|---|---|
| Sample: | Post-period | Post-period |
| Dependent variable: | Offering supply? | Offering supply? |
| Percent pre-period days open | 0.7073*** | 0.6968*** |
| | *(0.0082)* | *(0.0087)* |
| Network measure of competition | 0.4795*** | 0.4016*** |
| | *(0.0135)* | *(0.0275)* |
| Average meal price | -0.0015*** | -0.0014*** |
| | *(0.0002)* | *(0.0002)* |
| Eater rating | 0.0121*** | 0.0097*** |
| | *(0.0016)* | *(0.0016)* |
| Uber age | 0.0435*** | 0.0437*** |
| | *(0.0011)* | *(0.0009)* |
| Log mean household income | -0.0632*** | 0.0013 |
| | *(0.0024)* | *(0.0043)* |
| Smartphone penetration | -0.0814*** | 0.2978*** |
| | *(0.0118)* | *(0.0427)* |
| Broadband penetration | 0.1585*** | -0.2361*** |
| | *(0.0094)* | *(0.0316)* |
| Log population density | -0.0278*** | -0.0176*** |
| | *(0.0006)* | *(0.0011)* |
| | | |
| City fixed effects? | Yes | No |
| ZIP code fixed effects? | No | Yes |
| Date fixed effects? | Yes | Yes |
| | | |
| R-squared | 15.7% | 19.4% |
| Restaurant-day observations | 1,033,123 | 1,015,805 |

Note: This analysis considers what restaurant- and locality-level characteristics are related to the likelihood that the restaurant offers supply in the post-period. The dependent variable is a binary variable that takes the value of 1 if a restaurant offers supply on a given day during the post-period. Standard errors are clustered by date. ***, **, and * represents significance at $p < 0.001$, 0.01, and 0.05, respectively.

**Appendix Table A5: Regression Estimate of the Change in Daily Restaurant Orders on the Uber Eats Platform in Response to the COVID-19 Pandemic with an Alternative Sample**

| | (1) | (2) | (3) |
|---|---|---|---|
| Sample: | All restaurant-days excluding those with no supply time pre-shock | Restaurant-days with supply excluding those with no supply time pre-shock | Restaurant-days with supply excluding those with no supply time pre-shock |
| Dependent variable: | Daily orders | Daily orders | Daily orders |
| Post | 0.1149*** | 0.4122*** | 0.4824*** |
| | *(0.0078)* | *(0.0086)* | *(0.0210)* |
| Post x Dallas | | | 0.1588*** |
| | | | *(0.0312)* |
| Post x Miami | | | -0.2017*** |
| | | | *(0.0279)* |
| Post x New York City | | | 0.0729* |
| | | | *(0.0307)* |
| Post x San Francisco | | | 0.2359*** |
| | | | *(0.0348)* |
| Restaurant fixed effects? | Yes | Yes | Yes |
| Day of week fixed effects? | Yes | Yes | Yes |
| R-squared | 73.9% | 79.3% | 79.5% |
| Restaurant-day observations | 2,568,938 | 1,796,075 | 1,796,075 |

Note: This analysis is analogous to Table 2 in the draft except the sample only includes restaurants that offer supply during some point of the pre-period. ***, **, and * represents significance at $p < 0.001$, 0.01, and 0.05, respectively.

## Appendix Table A6: Regression Estimate of the Effect of Supply on Total Orders in a Locality

| | (1) | (2) | (3) | (4) | (5) | (6) | (7) | (8) | (9) |
|---|---|---|---|---|---|---|---|---|---|
| Sample: | All sample | All sample | All sample | Pre-period | Pre-period | Pre-period | Post-period | Post-period | Post-period |
| Dependent variable: | Daily orders | Daily orders | Daily orders | Daily orders | Daily orders | Daily orders | Daily orders | Daily orders | Daily orders |
| Restaurants open | 2.285*** | | | 1.649*** | | | 3.548*** | | |
| | *(0.3284)* | | | *(0.3165)* | | | *(0.5273)* | | |
| Log restaurants open | | 40.27*** | | | 28.57*** | | | 45.03*** | |
| | | *(5.5580)* | | | *(6.2910)* | | | *(9.1090)* | |
| Percent restaurants open | | | 64.76*** | | | 43.94*** | | | 75.93*** |
| | | | *(8.4890)* | | | *(7.2830)* | | | *(13.2200)* |
| ZIP code fixed effects? | Yes | Yes | Yes | Yes | Yes | Yes | Yes | Yes | Yes |
| Date fixed effects? | Yes | Yes | Yes | Yes | Yes | Yes | Yes | Yes | Yes |
| R-squared | 93.8% | 93.5% | 93.5% | 96.6% | 96.6% | 96.6% | 94.9% | 94.7% | 94.6% |
| ZIP code-day observations | 86,743 | 86,743 | 86,743 | 46,744 | 46,744 | 46,744 | 39,999 | 39,999 | 39,999 |

Note: The table presents regression results that assess the relationship between restaurant supply on the Uber Eats platform and total demand on the platform. This analysis is analogous to Table 4, except it uses the raw count of orders at the ZIP code level as a dependent variable rather than the log-transformed count of orders. The data set is a balanced panel consisting of ZIP code-days. The dependent variable is the log-transformed count of orders on the Uber Eats platform to restaurants located within the ZIP code. Standard errors are clustered at the ZIP code-level. ***, **, and * represents significance at $p < 0.001$, 0.01, and 0.05, respectively.

**Appendix Table A7: Temporal Effects of the Post-Period**

| | (1) | (2) |
|---|---|---|
| Sample: | All restaurant-days | Restaurant-days with supply |
| Dependent variable: | Daily orders | Daily orders |
| Time from Post | | |
| Eight weeks before | 0.0383*** | 0.0031 |
| | *(0.0084)* | *(0.0076)* |
| Seven weeks before | 0.0703*** | 0.0275*** |
| | *(0.0058)* | *(0.0054)* |
| Six weeks before | 0.0535*** | 0.0073 |
| | *(0.0048)* | *(0.0046)* |
| Five weeks before | 0.0549*** | 0.0078. |
| | *(0.0048)* | *(0.0045)* |
| Four weeks before | 0.0373*** | -0.009* |
| | *(0.0047)* | *(0.0044)* |
| Three weeks before | 0.0485*** | -0.0016 |
| | *(0.0044)* | *(0.0043)* |
| Two weeks before | -0.0347*** | -0.0814*** |
| | *(0.0040)* | *(0.0040)* |
| Week of shock | 0.0784*** | 0.2192*** |
| | *(0.0056)* | *(0.0065)* |
| Two weeks after shock | 0.1074*** | 0.3674*** |
| | *(0.0069)* | *(0.0082)* |
| Three weeks after shock | 0.1274*** | 0.4730*** |
| | *(0.0080)* | *(0.0095)* |
| Four weeks after shock | 0.2070*** | 0.5969*** |
| | *(0.0089)* | *(0.0111)* |
| Five weeks after shock | 0.1918*** | 0.5175*** |
| | *(0.0086)* | *(0.0107)* |
| Six weeks after shock | 0.2076*** | 0.4956*** |
| | *(0.0090)* | *(0.0110)* |
| Seven weeks after shock | 0.2705*** | 0.5833*** |
| | *(0.0131)* | *(0.0162)* |
| Restaurant fixed effects? | Yes | Yes |
| Day of week fixed effects? | Yes | Yes |
| R-squared | 73.9% | 79.4% |
| Restaurant-day observations | 2,727,709 | 1,862,748 |

Note: The table presents regression results that present that show how the increase in orders manifests temporally during the post-period. For this regression, the post-period indicator variable is replaced with a categorical variable measuring the distance in time from the imposition of the shelter-in-place orders in weeks. For this regression, the omitted group is the week prior to the shock. ***, **, and * represents significance at $p < 0.001$, 0.01, and 0.05, respectively.